\documentclass[preprint,pra,showpacs,showkeys,tightenlines]{revtex4}

\usepackage{graphicx}% Include figure files
\usepackage{dcolumn}% Align table columns on decimal point
\usepackage{bm}% bold math
\usepackage{feynmp}
\usepackage{nicefrac}

\def\half{\nicefrac{1}{2}}
\def\etal{\emph{et~al.}}

\def\cm{cm$^{-1}$}
\newcommand{\eref}[1]{(\ref{#1})}
\newcommand{\Eref}[1]{Eq.~(\ref{#1})}
\newcommand{\Tref}[1]{Table~\ref{#1}}
\newcommand{\Cite}[1]{Ref.~\cite{#1}}
\newcommand{\Fig}[1]{Fig.~\ref{#1}}
\newcommand{\vect}[1]{\bm{#1}}
\newcommand{\scaps}[1]{$\,${\footnotesize #1}}
\newcommand{\threej}[3]{\left( \begin{array}{ccc} #1 & #2 & #3 \\ \half & -\half & 0 \end{array} \right) }
\newcommand{\sixj}[6]{\left\{ \begin{array}{ccc} #1 & #2 & #3 \\ #4 & #5 & #6 \end{array} \right\} }
\begin{document}

\title{Calculation of isotope shifts and relativistic shifts in CI, CII, CIII and CIV}

\author{J. C. Berengut}
\email{jcb@phys.unsw.edu.au}
\affiliation{School of Physics, University of New South Wales, Sydney 2052, Australia}
\author{V. V. Flambaum}
\affiliation{School of Physics, University of New South Wales, Sydney 2052, Australia}
\affiliation{Physics Division, Argonne National Laboratory, Argonne, Illinois 60439-4843, USA}
\author{M. G. Kozlov}
\affiliation{Petersburg Nuclear Physics Institute, Gatchina, 188300, Russia}

\date{\today}

\begin{abstract}

We present an accurate \emph{ab~initio} method of calculating isotope shifts and relativistic shifts in atomic spectra. We test the method on neutral carbon and three carbon ions. The relativistic shift of carbon lines may allow them to be included in analyses of quasar absorption spectra that seek to measure possible variations in the fine structure constant $\alpha$ over the lifetime of the Universe. Carbon isotope shifts can be used to measure isotope abundances in gas clouds: isotope abundances are potentially an important source of systematic error in the $\alpha$--variation studies. These abundances are also needed to study nuclear reactions in stars and supernovae, and test models of chemical evolution of the Universe.

\end{abstract}

\pacs{31.30.Gs, 06.20.-f, 31.25.Jf}
\keywords{isotope shift; mass shift; relativistic shift; carbon}

\maketitle

\section{\label{sec:intro} Introduction}

Recent studies of quasar absorption systems have suggested that the fine structure constant, $\alpha = e^2/\hbar c$, may have been smaller in the early universe than it is today on Earth \cite{webb99prl,webb01prl,murphy01mnrasA,webb03ass}. Other groups using different telescopes have shown zero variation \cite{quast04aap,srianand04prl,levshakov05aap}. All of these studies use the ``many multiplet'' method \cite{dzuba99prl}, which relies on the comparison of the wavelengths of many transitions in many ions to enhance the size of the effects and remove sources of systematic error. While this method offers an order-of-magnitude improvement in sensitivity over the previously used alkali-doublet method, a potential systematic error is introduced related to the isotope abundances of the absorbers. If the isotope abundances of the absorbers differ from the terrestrial abundances then there can be spurious shifts in the measured wavelengths which may be incorrectly interpreted as variation of $\alpha$.

This problem can be resolved if both the relativistic shift and isotope shift of each transition is known, by using particular combinations of the transition frequencies as probes which are insensitive to either $\alpha$-variation or isotopic abundances \cite{kozlov04pra}. Changes in the isotope abundances and the fine-structure constant can then be measured directly.

The measured isotopic abundance of carbon can be used to test models of chemical evolution of the Universe. Of particular importance are some chemical evolution models with enhanced populations of intermediate-mass stars that serve as factories for heavy Mg isotopes. The changes in the relative abundances of Mg isotopes could account for much of the evidence for variation in $\alpha$ at relatively low redshifts ($z \lesssim 2$) \cite{ashenfelter04prl,ashenfelter04apj}. However, these models also overproduce nitrogen, violating observed abundance trends in high-$z$ damped Lyman-$\alpha$ systems. Furthermore, it has been shown that such models would also increase the ratio of $^{13}$C to $^{12}$C \cite{fenner05mnras}. With the isotope shifts calculated in this paper it is possible to measure this ratio, and hence provide a diagnostic of these non-standard chemical evolution models.

We also have one more reason to study carbon: it is a well studied atom, and we can compare the results of our method with those of other theoretical analyses, as well as a few experiments (unfortunately, isotope shifts for the majority of lines used in astrophysical applications have not been measured). In particular, much progress has been made to calculate isotope shifts using the multiconfigurational Hartree-Fock (MCHF) and configuration interaction (CI) approach \cite{carlsson95jpb,jonsson96jpb,jonsson99jpb,godefroid01jpb}. In Sections~\ref{sec:method} and \ref{sec:energy} we present our method of calculating the isotope shift. It uses the finite-field method to reduce the problem of isotope shift (or relativistic shift) to that of an energy calculation. The transition energies are calculated using a combination of the configuration interaction and many-body perturbation theory (MBPT) approaches, following the works of Refs.~\cite{dzuba96pra} and~\cite{dzuba98pra}.

In Sec.~\ref{sec:calc} we show that our CI + MBPT method has accuracy comparable to that of the MCHF calculations, and are accurate enough to be used to measure isotope abundances. We believe that our method may have some advantages over the MCHF calculations; in particular, it is more readily applicable to heavier atoms (where we need to calculate isotope shifts for the study of the $\alpha$ variation and isotopic evolution) and has already been shown to be more accurate in the case of Mg\scaps{I} \cite{berengut05arXiv}. We also use this method to calculate dependence of the carbon transition frequencies on $\alpha$, which is needed to study $\alpha$ variation.

\section{\label{sec:method} Method}

Using many-body perturbation theory in the residual Coulomb operator and specific mass shift (SMS) operator to calculate isotope shift shows poor convergence. Therefore, we are looking for an ``all order'' method of calculation. The finite-field scaling method is used, reducing the task to an energy calculation, and including the SMS in all parts of the calculation. A similar idea is used to calculate relativistic shift.

\subsection{Relativistic shift}
To measure $\alpha$ in the distant past, we compare the frequency of transitions in the laboratory, $\omega_0$, with those measured in the quasar absorption spectra, $\omega$. The difference can be expressed as
\begin{equation}
    \omega = \omega_0 + q x,
\end{equation}
where $x = (\alpha / \alpha_0 )^2 - 1$ and $\alpha_0$ is the laboratory value of the fine structure constant. We vary $\alpha$ directly in computer codes, and extract the relativistic shift $q$ from the calculated spectrum as
\begin{equation}
    q = \frac{d \omega}{d x} \bigg{|}_{x = 0} .
\end{equation}
Thus we have reduced the problem to an accurate numerical calculation of the energy, and hence $\omega$.

\subsection{Isotope shift}

The isotope shifts of atomic transition frequencies come from two sources: the finite size of the nuclear charge distribution (the ``volume'' or ``field'' shift), and the finite mass of the nucleus (see, e.g., \Cite{sobelman79book}). In atoms as light as carbon, the field shift is negligible in comparison to the mass shift: we will not consider it further.

Because a real nucleus has a finite mass, there is a recoil effect from the movements of the electrons.
The energy shift due to the recoil of the nucleus of mass $M$ is given by
\begin{eqnarray}
\frac{\vect{p}_N^2}{2M} &=& \frac{1}{2M} \left( \sum_i \vect{p}_i \right)^2 \nonumber \\
     &=& \frac{1}{2M} \sum_i p_i^2 + \frac{1}{M} \sum_{i<j} \vect{p}_i \cdot \vect{p}_j . \label{eq:smsdef}
\end{eqnarray}
This ``mass shift'' is traditionally divided into the normal mass shift (NMS) and the specific mass shift (SMS),
given by the first and second terms of \Eref{eq:smsdef}, respectively.
The normal mass shift is easily calculated from the transition frequency; the specific mass shift
is difficult to evaluate accurately.

The shift in energy of any transition in an isotope with mass number
$A'$ with respect to an isotope with mass number $A$ can be expressed
as
\begin{equation}
\label{eq:is}
\delta \nu^{A', A}
    = \nu^{A'} - \nu^{A}
    = \left( k_{\rm NMS} + k_{\rm SMS} \right)\left(\frac{1}{A'} - \frac{1}{A} \right),
\end{equation}
where the normal mass shift constant is
\begin{equation}
    k_{\rm NMS} = -\frac{\nu}{1823} .
\end{equation}
The value 1823 refers to the ratio of the atomic mass unit to the electron mass.

To calculate $k_{\rm SMS}$ we include a scaled specific-mass-shift operator directly into our energy calculation from the very beginning. We add the two-body SMS operator to the Coulomb potential
$\tilde{Q} = 1/\left| \vect{r}_1 - \vect{r}_2 \right| + \lambda \vect{p}_1 \cdot \vect{p}_2$. The operator $\tilde{Q}$ replaces the Coulomb operator everywhere that it appears in an energy calculation. We recover the specific mass shift constant as
\begin{equation}
    k_{\rm SMS} = \frac{d \omega}{d \lambda} \bigg{|}_{\lambda = 0}.
\end{equation}
The operator $\tilde{Q}$ has the same symmetry and structure as the Coulomb operator (see Appendix~\ref{app:sms_operator}). 

\section{\label{sec:energy} Energy calculation}

To calculate the energies we first solve the Dirac-Fock equations for the core and valence electrons, and we generate a basis set that includes the core and valence orbitals and a number of virtual orbitals. We then calculate the energy levels using a combination of CI and MBPT for many-valence-electron atoms as was done for Mg\scaps{I} in \Cite{berengut05arXiv}. In this section we outline the procedure for the combined CI and MBPT method; it generally follows the work of \Cite{dzuba96pra}. Note that for single-valence-electron atoms, where the CI procedure is unnecessary, the method reduces to the addition of core-correlation effects to the Dirac-Fock energy using MBPT, as was shown to be highly successful for calculating SMS in \Cite{berengut03pra}. Atomic units ($\hbar = m_e = e = 1$) are used throughout this paper.

\subsection{Single particle basis}

We firstly solve the Dirac-Fock equations for one-particle wavefunctions of the open-shell core $\left| m \right>$
\begin{equation}
\label{eq:single_particle}
h^{\rm DF} \left| m \right> = \epsilon_m \left| m \right> ,
\end{equation}
where
\begin{equation}
h^{\rm DF} = c\, \vect{\alpha} \cdot \vect{p} + (\beta -1) m_e c^2 - \frac{Z}{r} + V^{\rm N_{DF}}(r)
\end{equation}
where $V^{\rm N_{DF}}$ is the potential (both direct and exchange) of the $N_{\rm DF}$ electrons included in the self-consistent Hartree-Fock procedure. Note that this is not necessarily the number of electrons in the closed-shell core, in fact $N_{\rm core} \leq N_{\rm DF} \leq N$, where $N$ is the total number of electrons. For the purposes of the CI calculation there are $N - N_{\rm core}$ valence electrons.

We need to generate a basis set, $\left| i \right>$ that includes the core and valence states and a large number of virtual states. In this paper we have used a B-spline basis set, formed by diagonalizing the Dirac-Fock operator on the basis set of B-splines and excluding orbitals with high energy. This basis has been shown to be effective in energy calculations using this method of CI and MBPT \cite{dzuba98pra}.

\subsection{Configuration interaction method}

The many-electron Hilbert space is separated into subspaces $P$ and $Q$. $P$ corresponds to the frozen-core approximation; $Q$ is complimentary to it and includes all states with core excitations. Using Slater determinants (denoted with capital letters) $\left| I \right>$ of the single-particle functions $\left| i \right>$ as a basis set in the many-electron space, we can define projection operators for $P$ and $Q$ by
\begin{eqnarray}
\mathcal{P} = \sum_{I \in P} \left| I \right> \left< I \right| \\
\mathcal{P} + \mathcal{Q} = 1
\end{eqnarray}
Determinants that have all core states fully occupied are in the $P$ subspace; all others are in the subspace $Q$.

In the CI method, the many-electron wavefunction is expressed as a linear combination of Slater determinants from the subspace $P$:
\begin{equation}
\psi = \sum_{I \in P} C_I \left| I \right> ,
\end{equation}
where the $C_I$ are obtained from the matrix eigenvalue problem
\begin{equation}
\label{eq:CI_matrix}
\sum_{J \in P} H_{IJ} C_J = E C_I .
\end{equation}
In the frozen-core approximation, the determinants $\left| I \right>$ need include only the valence electrons. Although $P$ is infinite-dimensional, we can use a finite-dimensional model subspace by specifying the set of allowed configurations for the valence electrons, for example by restricting the set of single particle states. The restrictions we use are different for each ion, and are expressed more fully in the relevant parts of Section~\ref{sec:calc}. We will not distinguish here between the finite and infinite subspaces.

The Hamiltonian for the CI problem is a projection of the exact Hamiltonian $\mathcal{H}$ onto the model subspace. The core is frozen in the $P$ subspace, so our projected Hamiltonian is
\begin{equation}
\label{eq:PHP}
\mathcal{PHP} = E_{\rm core} + \sum_{i > N_{\rm core}} h^{\rm CI}_i + \sum_{j > i > N_{\rm core}} \frac{1}{r_{ij}}
\end{equation}
where $E_{\rm core}$ is the total energy of the $N_{\rm core}$ core electrons and the single particle operator
\begin{equation}
\label{eq:h_CI}
h^{\rm CI} = c\, \vect{\alpha} \cdot \vect{p} + (\beta - 1) m_e c^2 -\frac{Z}{r} + V^{\rm N_{core}}
\end{equation}
acts only on the valence electrons. Using the operator~\eref{eq:PHP} in \Eref{eq:CI_matrix} corresponds to the pure CI method in the frozen-core approximation.

\subsection{Exact Hamiltonian expansion}

We wish to write the exact equivalent of $\mathcal{H}$ in the subspace $P$. The ``Feshbach operator'' yields the exact energy when operating on the model function $\Psi_P = \mathcal{P} \Psi$. Following Lindgren and Morrison \cite{lindgren86book}, we start from the many-body Schr\"odinger equation in the form
\begin{equation}
\label{eq:schrodinger}
\mathcal{H(P+Q)} \Psi = E \Psi
\end{equation}
and operate from the left with $\mathcal{P}$ and $\mathcal{Q}$, respectively to obtain
\begin{eqnarray*}
\mathcal{PHP}\, \Psi_P + \mathcal{PHQ}\, \Psi_Q &=& E \Psi_P \\
\mathcal{QHP}\, \Psi_P + \mathcal{QHQ}\, \Psi_Q &=& E \Psi_Q .
\end{eqnarray*}
Eliminating $\Psi_Q$,
\begin{equation}
\label{eq:exact_expansion}
\left[ \mathcal{PHP} + \Sigma (E) \right] \Psi_P = E \Psi_P
\end{equation}
where
\begin{equation}
\label{eq:sigma_def}
\Sigma (E) = \mathcal{PHQ} \frac{1}{E - \mathcal{QHQ}} \mathcal{QHP} .
\end{equation}

In \Cite{dzuba96pra} these expressions were also used to rewrite the orthonormality conditions for the solutions of \Eref{eq:schrodinger} in terms of the model functions $\Psi_P$ (the solutions of \Eref{eq:exact_expansion}). As they found however, if an appropriate choice of the $P$ subspace is made, the usual orthonormality condition for $\Psi_P$ can be applied. In this case the standard CI procedure can be used to solve \Eref{eq:exact_expansion}.

\subsection{Many-body perturbation theory}

Here we will generate a perturbation expansion for $\Sigma$. Define a single particle Hamiltonian by
\begin{eqnarray}
h_0 a_i^\dag \left| 0 \right> &=& \epsilon_i a_i^\dag \left| 0 \right> \\
\epsilon_i & \equiv & \left< i \right| h^{\rm DF} \left| i \right> . \nonumber
\end{eqnarray}
where we introduce the operators $a^\dag_i$ ($a_i$) to create (annihilate) a particle. For particles in the core, the functions are those of \Eref{eq:single_particle}. The many-body zero-order Hamiltonian is
\begin{equation}
\mathcal{H}_0 = E_{\rm core} + \sum_i \{ a_i^\dag a_i \} \epsilon_i .
\end{equation}
%\begin{eqnarray}
%\mathcal{H}_0 &=& \sum_i a_i^\dag a_i \epsilon_i \nonumber \\
%              &=& \sum_m^{\rm core} \epsilon_m + \sum_i \{ a_i^\dag a_i \} \epsilon_i .
%\end{eqnarray}
where the brackets $\{...\}$ denote normal ordering with respect to the closed-shell core.

The exact Hamiltonian is
\begin{equation}
\mathcal{H} = \sum_i h^{\rm nuc} + \sum_{i < j} \frac {1}{r_{ij}}
\end{equation}
where $h^{\rm nuc} = c\,\vect{\alpha} \cdot \vect{p} + (\beta-1) m_e c^2 - \frac{Z}{r_i}$. $\mathcal{H}$ can be separated into zero, one, and two-body parts:
\begin{eqnarray}
%\mathcal{H}^{(0)} &=& \sum_m^{\rm core} h^{\rm nuc}    + \frac{1}{2} \sum_{mn}^{\rm core}
%           \left( \left< mn | r_{12}^{-1} | mn \right> - \left< mn | r_{12}^{-1} | nm \right> \right) \\
\mathcal{H}^{(0)} &=& E_{\rm core} \nonumber \\
\mathcal{H}^{(1)} &=& \sum_{ij} \{ a_i^\dag a_j \}
        \bigg[ \left< i \right| h^{\rm nuc} \left| j \right> \nonumber \\
    & & + \sum_{m}^{\rm core} \left( \left< im \right| r_{12}^{-1} \left| jm \right>
                                   - \left< im \right| r_{12}^{-1} \left| mj \right>
                              \right) \bigg] \nonumber \\
\label{eq:H_1}
    &=& \sum_{ij} \{ a_i^\dag a_j \} \left< i \right| h^{\rm CI} \left| j \right> \\
\label{eq:H_2}
\mathcal{H}^{(2)} &=& \sum_{ijkl} \{ a_i^\dag a_j^\dag a_l a_k \} \left< ij \right| r_{12}^{-1} \left| kl \right> .
\end{eqnarray}

Expanding \Eref{eq:sigma_def} in the residual Coulomb interaction, $\mathcal{V} = \mathcal{H} - \mathcal{H}_0$, we obtain
\begin{eqnarray}
\Sigma (E) &=& \mathcal{PHQ}\frac{1}{E - \mathcal{H}_0}\mathcal{QHP} \nonumber \\
           & & + \mathcal{PHQ}\frac{1}{E - \mathcal{H}_0}\mathcal{QVQ}\frac{1}{E - \mathcal{H}_0}\mathcal{QHP}
               + \ldots
\label{eq:sigma_expansion}
\end{eqnarray}

One advantage of this formalism is that $h_0$ is not necessarily the same as $h^{\rm DF}$. Thus we may in principle use any set of functions we like in the virtual basis as long as they are orthogonal. In practice, however, it is important that $\mathcal{V}$ not get too large, and this requires that
\begin{equation}
\label{eq:V_1}
\mathcal{V}^{(1)} = \left< i \right| h^{\rm CI}-h_0 \left| j \right>
% = \left< i \right| h^{\rm nuc}-h_0 \left| j \right> 
% + \sum_m^{\rm core} \left< im \right| r_{12}^{-1} \left| jm \right> -
%                     \left< im \right| r_{12}^{-1} \left| mj \right>
\end{equation}
is small.

We can write $\Sigma$ in matrix form:
\begin{eqnarray}
\Sigma_{IJ} &=& \sum_{M \in Q} \frac{\left< I \right| H \left| M \right> \left< M \right| H \left| J \right>}
                                    {E - E_{M}} \nonumber \\
    & & + \sum_{M,L \in Q} \frac{\left< I \right| H \left| M \right> \left< M \right| V \left| L \right> 
                               \left< L \right| H \left| J \right>}{(E - E_M)(E - E_L)} + \ldots \label{eq:sigma_matrix} \\
    &=& \left( \Sigma_2 \right)_{IJ} + \left( \Sigma_3 \right)_{IJ} + ...
\end{eqnarray}
where I and J enumerate determinants from the subspace $P$, and M and L are determinants from the subspace $Q$.

In this paper we calculate $\Sigma$ to second-order of the perturbation expansion. For the one-valence-electron case it has been shown that this level of perturbation theory, when used with the finite-field scaling method, is sufficient to obtain accurate results for isotope shift \cite{berengut03pra}. Furthermore, previous studies have shown that energies calculated to this order are very accurate \cite{dzuba96pra,dzuba98pra}.

Substituting $\Sigma_2$ into \eref{eq:exact_expansion} we obtain the equation of the combined CI and MBPT method, which we write in the same form as \Eref{eq:CI_matrix}:
\begin{equation}
\label{eq:CI_MBPT_matrix}
\sum_{J \in P} \left( H_{IJ} + 
     \sum_{M \in Q} \frac{\left< I \right| H \left| M \right> \left< M \right| H \left| J \right>}
                         {E - E_{M}}
     \right) C_J = E C_I .
\end{equation}
Thus our method includes the core-correlation effects by simply altering the matrix elements in the CI calculation. In practice, this involves adding the matrix element of the sigma operator to the one and two-particle Coulomb integrals.

\subsection{Diagrammatic technique for calculating $\Sigma$}

Here we present all second-order diagrams for $\Sigma$, represented by Goldstone diagrams (see, e.g.:~\Cite{lindgren86book}). The diagrams are valid for any ion and any choice of core. Unlinked lines are omitted, since they correspond to states of the valence electrons that are not involved in the interaction.  The omitted lines do affect the value of the MBPT correction via the Pauli principle; however, as noted in \Cite{dzuba96pra}, the Pauli principle can be ignored due to the exact cancellation of unphysical terms in different diagrams. This rule works for all orders of perturbation theory (this is Wick's theorem, see e.g.:~\Cite{lindgren86book}).

\begin{figure}[htb]
\caption{One-valence-electron diagrams of $\Sigma$ ($\Sigma^{(1)}$).}
\begin{fmffile}{sigma1}
\label{feyn:sigma1}
%MBPT 1.1
\fmfframe(10,20)(10,30){
\begin{fmfgraph*}(80,25)
  \fmfstraight
  \fmfpen{thin}
  \fmfset{arrow_len}{3mm}
  %external vertices
  \fmfleft{i2,i1}
  \fmfright{o2,o1}
  %top
  \fmffixedy{0}{i1,v1}
  \fmffixedy{0}{v1,v3}
  \fmffixedy{0}{v3,o1}
  \fmf{fermion,label=$a$,l.s=left,l.d=4}{i1,v1}
  \fmf{fermion,tension=0.5,label=$\beta$,l.s=left,l.d=4}{v1,v3}
  \fmf{fermion,label=$b$,l.s=left,l.d=4}{v3,o1}
  %bottom
  \fmffixedy{0}{i2,v2}
  \fmffixedy{0}{v2,v4}
  \fmffixedy{0}{v4,o2}
  \fmf{phantom}{i2,v2}
  \fmf{phantom}{v4,o2}
  \fmf{fermion,left=0.5,tension=0.25,label=$\alpha$,l.s=left,l.d=4}{v2,v4}
  \fmf{fermion,left=0.5,tension=0.25,label=$n$,l.s=right,l.d=4}{v4,v2}
  %vertical
  \fmf{dashes}{v1,v2}
  \fmf{dashes}{v3,v4}
  %diagram label
  \fmfforce{(0.5w,-25)}{c}
  \fmfv{label=$1$,l.d=0}{c}
\end{fmfgraph*}
}
%MBPT 1.2
\fmfframe(10,20)(10,30){
\begin{fmfgraph*}(80,25)
  \fmfstraight
  \fmfpen{thin}
  \fmfset{arrow_len}{3mm}
  %external vertices
  \fmfleft{i2,i1}
  \fmfright{o2,o1}
  %top
  \fmffixedy{0}{i1,v1}
  \fmffixedy{0}{v1,v3}
  \fmffixedy{0}{v3,o1}
  \fmf{fermion,label=$a$,l.s=left,l.d=4}{i1,v1}
  \fmf{fermion,tension=0.5,label=$\beta$,l.s=left,l.d=4}{v1,v3}
  \fmf{phantom}{v3,o1}
  %bottom
  \fmffixedy{0}{i2,v2}
  \fmffixedy{0}{v2,v4}
  \fmffixedy{0}{v4,o2}
  \fmf{phantom}{i2,v2}
  \fmf{fermion,label=$b$,l.s=left,l.d=4}{v4,o2}
  \fmf{fermion,tension=0.5,label=$\alpha$,l.s=right,l.d=4}{v2,v4}
  \fmf{fermion,tension=0,label=$n$,l.s=right,l.d=3}{v3,v2}
  %vertical
  \fmf{dashes}{v1,v2}
  \fmf{dashes}{v3,v4}
  %diagram label
  \fmfforce{(0.5w,-25)}{c}
  \fmfv{label=$2$,l.d=0}{c}
\end{fmfgraph*}
}
%MBPT 1.3
\fmfframe(10,20)(10,30){
\begin{fmfgraph*}(80,50)
  \fmfstraight
  \fmfpen{thin}
  \fmfset{arrow_len}{3mm}
  %external vertices
  \fmfleft{i3,i2,i1}
  \fmfright{o3,o2,o1}
  %top
  \fmffixedy{0}{i1,v1}
  \fmffixedy{0}{v1,v2}
  \fmffixedy{0}{v2,o1}
  \fmf{fermion,label=$a$,l.s=left,l.d=4}{i1,v1}
  \fmf{plain,tension=0.5}{v1,v2}
  \fmf{phantom}{v2,o1}
  \fmf{fermion,tension=0,label=$m$,l.s=right,l.d=3}{v2,v3}
  %middle
  \fmffixedy{0}{i2,v3}
  \fmffixedy{0}{v3,v4}
  \fmffixedy{0}{v4,o2}
  \fmf{phantom}{i2,v3}
  \fmf{plain,tension=0.5}{v3,v4}
  \fmf{fermion,label=$b$,l.s=left,l.d=4}{v4,o2}
  %bottom
  \fmffixedy{0}{i3,v5}
  \fmffixedy{0}{v5,v6}
  \fmffixedy{0}{v6,o3}
  \fmf{phantom}{i3,v5}
  \fmf{phantom}{v6,o3}
  \fmf{fermion,left=0.5,tension=0.25,label=$\alpha$,l.s=left,l.d=4}{v5,v6}
  \fmf{fermion,left=0.5,tension=0.25,label=$n$,l.s=right,l.d=4}{v6,v5}
  %vertical
  \fmf{dashes}{v3,v5}
  \fmf{dashes}{v2,v6}
  %diagram label
  \fmfforce{(0.5w,-25)}{c}
  \fmfv{label=$3$,l.d=0}{c}
\end{fmfgraph*}
}
%MBPT 1.4
\fmfframe(10,20)(10,30){
\begin{fmfgraph*}(80,50)
  \fmfstraight
  \fmfpen{thin}
  \fmfset{arrow_len}{3mm}
  %external vertices
  \fmfleft{i3,i2,i1}
  \fmfright{o3,o2,o1}
  %top
  \fmffixedy{0}{i1,v1}
  \fmffixedy{0}{v1,v2}
  \fmffixedy{0}{v2,o1}
  \fmf{fermion,label=$a$,l.s=left,l.d=4}{i1,v1}
  \fmf{plain,tension=0.5}{v1,v2}
  \fmf{phantom}{v2,o1}
  \fmf{fermion,tension=0,label=$m$,l.s=right,l.d=3}{v2,v3}
  %middle
  \fmffixedy{0}{i2,v3}
  \fmffixedy{0}{v3,v4}
  \fmffixedy{0}{v4,o2}
  \fmf{phantom}{i2,v3}
  \fmf{fermion,tension=0.5,label=$\alpha$,l.s=left,l.d=4}{v3,v4}
  \fmf{phantom}{v4,o2}
  \fmf{fermion,tension=0,label=$n$,l.s=left,l.d=3}{v4,v5}
  %bottom
  \fmffixedy{0}{i3,v5}
  \fmffixedy{0}{v5,v6}
  \fmffixedy{0}{v6,o3}
  \fmf{phantom}{i3,v5}
  \fmf{plain,tension=0.5}{v5,v6}
  \fmf{fermion,label=$b$,l.s=left,l.d=4}{v6,o3}
  %vertical
  \fmf{dashes}{v3,v5}
  \fmf{dashes}{v2,v4}
  %diagram label
  \fmfforce{(0.5w,-25)}{c}
  \fmfv{label=$4$,l.d=0}{c}
\end{fmfgraph*}
}
\end{fmffile}
\end{figure}
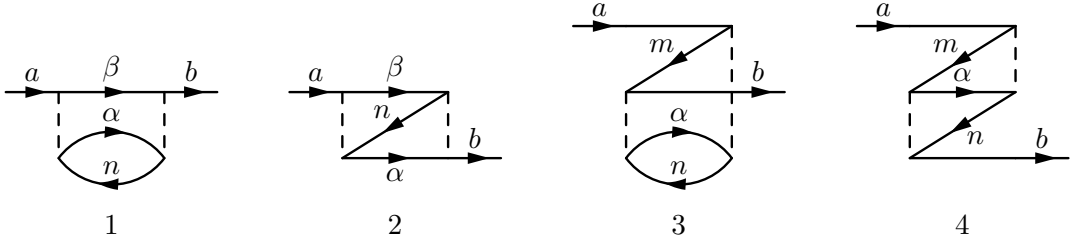

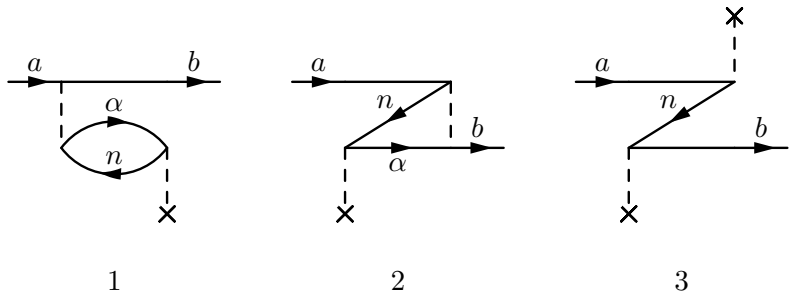
\begin{figure}[htb]
\caption{Subtraction diagrams of $\Sigma^{(1)}$. Diagrams 1 and 2 have complementary diagrams obtained by reflection in the vertical plane.}
\begin{fmffile}{sigma1sub}
\label{feyn:sigma1sub}
%SUB 1.1
\fmfframe(10,20)(10,30){
\begin{fmfgraph*}(80,50)
  \fmfstraight
  \fmfpen{thin}
  \fmfset{arrow_len}{3mm}
  %external vertices
  \fmfleft{i3,i2,i1}
  \fmfright{o3,o2,o1}
  %top
  \fmffixedy{0}{i1,v1}
  \fmffixedy{0}{v1,v2}
  \fmffixedy{0}{v2,o1}
  \fmf{fermion,label=$a$,l.s=left,l.d=4}{i1,v1}
  \fmf{plain,tension=0.5}{v1,v2}
  \fmf{fermion,label=$b$,l.s=left,l.d=4}{v2,o1}
  %middle
  \fmffixedy{0}{i2,v3}
  \fmffixedy{0}{v3,v4}
  \fmffixedy{0}{v4,o2}
  \fmf{phantom}{i2,v3}
  \fmf{fermion,left=0.5,tension=0.25,label=$\alpha$,l.s=left,l.d=4}{v3,v4}
  \fmf{fermion,left=0.5,tension=0.25,label=$n$,l.s=right,l.d=4}{v4,v3}
  \fmf{phantom}{v4,o2}
  %bottom
  \fmffixedy{0}{i3,v5}
  \fmffixedy{0}{v5,o3}
  \fmf{phantom,tension=1/3}{i3,v5}
  \fmf{phantom}{v5,o3}
  \fmfv{d.shape=cross,d.size=8}{v5}
  %vertical
  \fmf{dashes}{v4,v5}
  \fmf{dashes}{v1,v3}
  %diagram label
  \fmfforce{(0.5w,-25)}{c}
  \fmfv{label=$1$,l.d=0}{c}
\end{fmfgraph*}
}
%SUB 1.2
\fmfframe(10,20)(10,30){
\begin{fmfgraph*}(80,50)
  \fmfstraight
  \fmfpen{thin}
  \fmfset{arrow_len}{3mm}
  %external vertices
  \fmfleft{i3,i2,i1}
  \fmfright{o3,o2,o1}
  %top
  \fmffixedy{0}{i1,v1}
  \fmffixedy{0}{v1,v2}
  \fmffixedy{0}{v2,o1}
  \fmf{fermion,label=$a$,l.s=left,l.d=4}{i1,v1}
  \fmf{plain,tension=0.5}{v1,v2}
  \fmf{phantom}{v2,o1}
  \fmf{fermion,tension=0,label=$n$,l.s=right,l.d=3}{v2,v3}
  %middle
  \fmffixedy{0}{i2,v3}
  \fmffixedy{0}{v3,v4}
  \fmffixedy{0}{v4,o2}
  \fmf{phantom}{i2,v3}
  \fmf{fermion,tension=0.5,label=$\alpha$,l.s=right,l.d=4}{v3,v4}
  \fmf{fermion,label=$b$,l.s=left,l.d=4}{v4,o2}
  %bottom
  \fmffixedy{0}{i3,v5}
  \fmffixedy{0}{v5,o3}
  \fmf{phantom}{i3,v5}
  \fmf{phantom,tension=1/3}{v5,o3}
  \fmfv{d.shape=cross,d.size=8}{v5}
  %vertical
  \fmf{dashes}{v3,v5}
  \fmf{dashes}{v2,v4}
  %diagram label
  \fmfforce{(0.5w,-25)}{c}
  \fmfv{label=$2$,l.d=0}{c}
\end{fmfgraph*}
}
%SUB 1.3
\fmfframe(10,20)(10,30){
\begin{fmfgraph*}(80,75)
  \fmfstraight
  \fmfpen{thin}
  \fmfset{arrow_len}{3mm}
  %external vertices
  \fmfleft{i3,i2,i1,i0}
  \fmfright{o3,o2,o1,o0}
  %topmost
  \fmffixedy{0}{i0,v0}
  \fmffixedy{0}{v0,o0}
  \fmf{phantom,tension=1/3}{i0,v0}
  \fmf{phantom}{v0,o0}
  \fmfv{d.shape=cross,d.size=8}{v0}
  %top
  \fmffixedy{0}{i1,v1}
  \fmffixedy{0}{v1,v2}
  \fmffixedy{0}{v2,o1}
  \fmf{fermion,label=$a$,l.s=left,l.d=4}{i1,v1}
  \fmf{plain,tension=0.5}{v1,v2}
  \fmf{phantom}{v2,o1}
  \fmf{fermion,tension=0,label=$n$,l.s=right,l.d=3}{v2,v3}
  %middle
  \fmffixedy{0}{i2,v3}
  \fmffixedy{0}{v3,v4}
  \fmffixedy{0}{v4,o2}
  \fmf{phantom}{i2,v3}
  \fmf{plain,tension=0.5}{v3,v4}
  \fmf{fermion,label=$b$,l.s=left,l.d=4}{v4,o2}
  %bottom
  \fmffixedy{0}{i3,v5}
  \fmffixedy{0}{v5,o3}
  \fmf{phantom}{i3,v5}
  \fmf{phantom,tension=1/3}{v5,o3}
  \fmfv{d.shape=cross,d.size=8}{v5}
  %vertical
  \fmf{dashes}{v0,v2}
  \fmf{dashes}{v3,v5}
  %diagram label
  \fmfforce{(0.5w,-25)}{c}
  \fmfv{label=$3$,l.d=0}{c}
\end{fmfgraph*}
}
\end{fmffile}
\end{figure}

The second-order diagrams can be grouped according to how many external lines (valence electrons) they have. There are four one-valence-electron diagrams, all shown in \Fig{feyn:sigma1}. They correspond to the ``self-energy'' correction, which describes correlations between the valence electron and the core. Additionally, there are five so-called ``subtraction diagrams'' for the self-energy. These are diagrams that involve the external field $\mathcal{V}^{(1)}$ (\Eref{eq:V_1}), and are so named because in the Hartree-Fock field enters $\mathcal{V}^{(1)}$ with a minus sign.
In our formulation we instead use $\mathcal{H}^{(1)}$ as the external field (see Eqs.~\eref{eq:H_1}~and~\eref{eq:sigma_matrix}) which is equivalent when calculating $\Sigma_2$ because $\mathcal{PH}_0 \mathcal{Q} = 0$. Three of the $\Sigma^{(1)}$ subtraction diagrams are shown in \Fig{feyn:sigma1sub}; the other two are the mirror image partners of diagrams \ref{feyn:sigma1sub}.1 and \ref{feyn:sigma1sub}.2.

%--------------------------------------------------------------------------------------------

%--------------------------------------------------------------------------------------------

\begin{figure}[htb]
\caption{$\Sigma^{(2)}$ subtraction diagrams. Each has a complementary mirror-image partner.}
\begin{fmffile}{sigma2sub}
\label{feyn:sigma2sub}
%SUB 2.1
\fmfframe(10,20)(10,30){
\begin{fmfgraph*}(80,100)
  \fmfstraight
  \fmfpen{thin}
  \fmfset{arrow_len}{3mm}
  %external vertices
  \fmfleft{i4,i3,i2,i1,i0}
  \fmfright{o4,o3,o2,o1,o0}
  %topmost
  \fmffixedy{0}{i0,v0}
  \fmffixedy{0}{v0,o0}
  \fmf{phantom,tension=1/3}{i0,v0}
  \fmf{phantom}{v0,o0}
  \fmfv{d.shape=cross,d.size=8}{v0}
  %top
  \fmffixedy{0}{i1,v1}
  \fmffixedy{0}{v1,v2}
  \fmffixedy{0}{v2,o1}
  \fmf{fermion,label=$a$,l.s=left,l.d=4}{i1,v1}
  \fmf{plain,tension=0.5}{v1,v2}
  \fmf{phantom}{v2,o1}
  \fmf{fermion,tension=0,label=$n$,l.s=right,l.d=3}{v2,v3}
  %middle
  \fmffixedy{0}{i2,v3}
  \fmffixedy{0}{v3,v4}
  \fmffixedy{0}{v4,o2}
  \fmf{phantom}{i2,v3}
  \fmf{plain,tension=0.5}{v3,v4}
  \fmf{fermion,label=$c$,l.s=left,l.d=4}{v4,o2}
  %bottom
  \fmffixedy{0}{i3,v5}
  \fmffixedy{0}{v5,v6}
  \fmffixedy{0}{v5,o3}
  \fmf{fermion,label=$b$,l.s=left,l.d=4}{i3,v5}
  \fmf{plain,tension=0.5}{v5,v6}
  \fmf{fermion,label=$d$,l.s=left,l.d=4}{v6,o3}
  %vertical
  \fmf{dashes}{v0,v2}
  \fmf{dashes}{v3,v5}
  %diagram label
  \fmfforce{(0.5w,-25)}{c}
  \fmfv{label=$1$,l.d=0}{c}
\end{fmfgraph*}
}
%SUB 2.2
\fmfframe(10,20)(10,30){
\begin{fmfgraph*}(80,100)
  \fmfstraight
  \fmfpen{thin}
  \fmfset{arrow_len}{3mm}
  %external vertices
  \fmfleft{i4,i3,i2,i1,i0}
  \fmfright{o4,o3,o2,o1,o0}
  %top
  \fmffixedy{0}{i1,v1}
  \fmffixedy{0}{v1,v2}
  \fmffixedy{0}{v2,o1}
  \fmf{fermion,label=$a$,l.s=left,l.d=4}{i1,v1}
  \fmf{plain,tension=0.5}{v1,v2}
  \fmf{fermion,label=$c$,l.s=left,l.d=4}{v2,o1}
  %middle
  \fmffixedy{0}{i2,v3}
  \fmffixedy{0}{v3,v4}
  \fmffixedy{0}{v4,o2}
  \fmf{phantom}{i2,v3}
  \fmf{plain,tension=0.5}{v3,v4}
  \fmf{fermion,label=$d$,l.s=left,l.d=4}{v4,o2}
  \fmf{fermion,tension=0,label=$n$,l.s=left,l.d=3}{v6,v3}
  %bottom
  \fmffixedy{0}{i3,v5}
  \fmffixedy{0}{v5,v6}
  \fmffixedy{0}{v6,o3}
  \fmf{fermion,label=$b$,l.s=left,l.d=4}{i3,v5}
  \fmf{plain,tension=0.5}{v5,v6}
  \fmf{phantom}{v6,o3}
  %bottom
  \fmffixedy{0}{i4,v7}
  \fmffixedy{0}{v7,o4}
  \fmf{phantom,tension=1/3}{i4,v7}
  \fmf{phantom}{v7,o4}
  \fmfv{d.shape=cross,d.size=8}{v7}
  %vertical
  \fmf{dashes}{v1,v3}
  \fmf{dashes}{v6,v7}
  %diagram label
  \fmfforce{(0.5w,-25)}{c}
  \fmfv{label=$2$,l.d=0}{c}
\end{fmfgraph*}
}
\end{fmffile}
\end{figure}
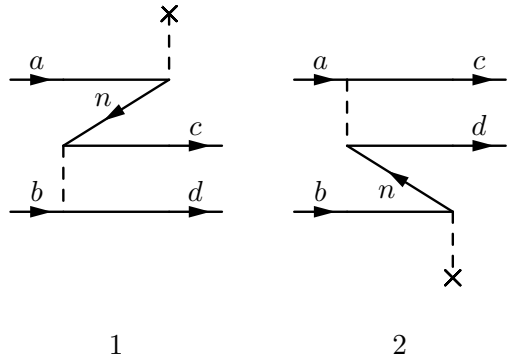

Diagrams with two external lines correspond to screening of the valence-valence interaction by the core electrons. There are nine diagrams represented in \Fig{feyn:sigma2}, and four subtraction diagrams represented in \Fig{feyn:sigma2sub}. The resulting corrections to the interactions between valence electrons can be written in the form of an effective radial integral, as is usually done for the Coulomb integrals. However, it is important to note that the box diagrams (\hbox{Figs. \ref{feyn:sigma2}.4 -- \ref{feyn:sigma2}.6}) have less symmetry than the Coulomb integrals, because swapping the initial and final states in either the upper or lower lines changes the integral. This approximately doubles the number of independent radial integrals that need to be stored for the CI problem. In addition, for the box diagrams the multipolarity of the Coulomb interaction need not follow the usual rule:
$\xi(l_a + l_c + k)\xi(l_b + l_d + k)$ (see Appendix~\ref{app:sms_operator}). Instead, $k$ need only satisfy the triangle conditions and can be both odd and even. This would again double the number of independent radial integrals, except that we have found that the diagrams of ``wrong'' parity are unimportant for carbon and may be omitted in order to reduce the complexity of the calculations.

%--------------------------------------------------------------------------------------------
\begin{figure}[htb]
\caption{$\Sigma^{(2)}$: Diagrams 1, 2 and 3 have complementary diagrams obtained by reflection in the vertical plane.}
\begin{fmffile}{sigma2}
\label{feyn:sigma2}
%MBPT 2.1
\fmfframe(10,20)(10,30){
\begin{fmfgraph*}(80,50)
  \fmfstraight
  \fmfpen{thin}
  \fmfset{arrow_len}{3mm}
  %external vertices
  \fmfleft{i3,i2,i1}
  \fmfright{o3,o2,o1}
  %top
  \fmffixedy{0}{i1,v1}
  \fmffixedy{0}{v1,v2}
  \fmffixedy{0}{v2,o1}
  \fmf{fermion,label=$a$,l.s=left,l.d=4}{i1,v1}
  \fmf{plain,tension=0.5}{v1,v2}
  \fmf{fermion,label=$c$,l.s=left,l.d=4}{v2,o1}
  %middle
  \fmffixedy{0}{i2,v3}
  \fmffixedy{0}{v3,v4}
  \fmffixedy{0}{v4,o2}
  \fmf{phantom}{i2,v3}
  \fmf{fermion,left=0.5,tension=0.25,label=$\alpha$,l.s=left,l.d=4}{v3,v4}
  \fmf{fermion,left=0.5,tension=0.25,label=$n$,l.s=right,l.d=4}{v4,v3}
  \fmf{phantom}{v4,o2}
  %bottom
  \fmffixedy{0}{i3,v5}
  \fmffixedy{0}{v5,v6}
  \fmffixedy{0}{v6,o3}
  \fmf{fermion,label=$b$,l.s=left,l.d=4}{i3,v5}
  \fmf{plain,tension=0.5}{v5,v6}
  \fmf{fermion,label=$d$,l.s=left,l.d=4}{v6,o3}
  %vertical
  \fmf{dashes}{v1,v3}
  \fmf{dashes}{v4,v6}
  %diagram label
  \fmfforce{(0.5w,-25)}{c}
  \fmfv{label=$1$,l.d=0}{c}
\end{fmfgraph*}
}
%MBPT 2.2
\fmfframe(10,20)(10,30){
\begin{fmfgraph*}(80,50)
  \fmfstraight
  \fmfpen{thin}
  \fmfset{arrow_len}{3mm}
  %external vertices
  \fmfleft{i3,i2,i1}
  \fmfright{o3,o2,o1}
  %top
  \fmffixedy{0}{i1,v1}
  \fmffixedy{0}{v1,v2}
  \fmffixedy{0}{v2,o1}
  \fmf{fermion,label=$a$,l.s=left,l.d=4}{i1,v1}
  \fmf{plain,tension=0.5}{v1,v2}
  \fmf{phantom}{v2,o1}
  \fmf{fermion,tension=0,label=$n$,l.s=right,l.d=3}{v2,v3}
  %middle
  \fmffixedy{0}{i2,v3}
  \fmffixedy{0}{v3,v4}
  \fmffixedy{0}{v4,o2}
  \fmf{phantom}{i2,v3}
  \fmf{fermion,tension=0.5,label=$\alpha$,label.s=right,l.d=4}{v3,v4}
  \fmf{fermion,label=$c$,l.s=left,l.d=4}{v4,o2}
  %bottom
  \fmffixedy{0}{i3,v5}
  \fmffixedy{0}{v5,v6}
  \fmffixedy{0}{v5,o3}
  \fmf{fermion,label=$b$,l.s=left,l.d=4}{i3,v5}
  \fmf{plain,tension=0.5}{v5,v6}
  \fmf{fermion,label=$d$,l.s=left,l.d=4}{v6,o3}
  %vertical
  \fmf{dashes}{v2,v4}
  \fmf{dashes}{v3,v5}
  %diagram label
  \fmfforce{(0.5w,-25)}{c}
  \fmfv{label=$2$,l.d=0}{c}
\end{fmfgraph*}
}
%MBPT 2.3
\fmfframe(10,20)(10,30){
\begin{fmfgraph*}(80,50)
  \fmfstraight
  \fmfpen{thin}
  \fmfset{arrow_len}{3mm}
  %external vertices
  \fmfleft{i3,i2,i1}
  \fmfright{o3,o2,o1}
  %top
  \fmffixedy{0}{i1,v1}
  \fmffixedy{0}{v1,v2}
  \fmffixedy{0}{v2,o1}
  \fmf{fermion,label=$a$,l.s=left,l.d=4}{i1,v1}
  \fmf{plain,tension=0.5}{v1,v2}
  \fmf{fermion,label=$c$,l.s=left,l.d=4}{v2,o1}
  %middle
  \fmffixedy{0}{i2,v3}
  \fmffixedy{0}{v3,v4}
  \fmffixedy{0}{v4,o2}
  \fmf{phantom}{i2,v3}
  \fmf{fermion,tension=0.5,label=$\alpha$,l.s=left,l.d=4}{v3,v4}
  \fmf{fermion,label=$d$,l.s=left,l.d=4}{v4,o2}
  \fmf{fermion,tension=0,label=$n$,l.s=left,l.d=3}{v6,v3}
  %bottom
  \fmffixedy{0}{i3,v5}
  \fmffixedy{0}{v5,v6}
  \fmffixedy{0}{v6,o3}
  \fmf{fermion,label=$b$,l.s=left,l.d=4}{i3,v5}
  \fmf{plain,tension=0.5}{v5,v6}
  \fmf{phantom}{v6,o3}
  %vertical
  \fmf{dashes}{v1,v3}
  \fmf{dashes}{v4,v6}
  %diagram label
  \fmfforce{(0.5w,-25)}{c}
  \fmfv{label=$3$,l.d=0}{c}
\end{fmfgraph*}
}
%MBPT 2.4
\fmfframe(10,20)(10,30){
\begin{fmfgraph*}(80,50)
  \fmfstraight
  \fmfpen{thin}
  \fmfset{arrow_len}{3mm}
  %external vertices
  \fmfleft{i3,i2,i1}
  \fmfright{o3,o2,o1}
  %top
  \fmffixedy{0}{i1,v1}
  \fmffixedy{0}{v1,v2}
  \fmffixedy{0}{v2,o1}
  \fmf{fermion,label=$a$,l.s=left,l.d=4}{i1,v1}
  \fmf{plain,tension=0.5}{v1,v2}
  \fmf{phantom}{v2,o1}
  \fmf{fermion,tension=0,label=$n$,l.s=right,l.d=3}{v2,v3}
  %middle
  \fmffixedy{0}{i2,v3}
  \fmffixedy{0}{v3,v4}
  \fmffixedy{0}{v4,o2}
  \fmf{phantom}{i2,v3}
  \fmf{plain,tension=0.5}{v3,v4}
  \fmf{fermion,label=$c$,l.s=left,l.d=4}{v4,o2}
  %bottom
  \fmffixedy{0}{i3,v5}
  \fmffixedy{0}{v5,v6}
  \fmffixedy{0}{v5,o3}
  \fmf{fermion,label=$b$,l.s=left,l.d=4}{i3,v5}
  \fmf{fermion,tension=0.5,label=$\alpha$,l.s=left,l.d=4}{v5,v6}
  \fmf{fermion,label=$d$,l.s=left,l.d=4}{v6,o3}
  %vertical
  \fmf{dashes}{v2,v6}
  \fmf{dashes}{v3,v5}
  %diagram label
  \fmfforce{(0.5w,-25)}{c}
  \fmfv{label=$4$,l.d=0}{c}
\end{fmfgraph*}
}
%MBPT 2.5
\fmfframe(10,20)(10,30){
\begin{fmfgraph*}(80,50)
  \fmfstraight
  \fmfpen{thin}
  \fmfset{arrow_len}{3mm}
  %external vertices
  \fmfleft{i3,i2,i1}
  \fmfright{o3,o2,o1}
  %top
  \fmffixedy{0}{i1,v1}
  \fmffixedy{0}{v1,v2}
  \fmffixedy{0}{v2,o1}
  \fmf{fermion,label=$a$,l.s=left,l.d=4}{i1,v1}
  \fmf{fermion,tension=0.5,label=$\alpha$,l.s=left,l.d=4}{v1,v2}
  \fmf{fermion,label=$c$,l.s=left,l.d=4}{v2,o1}
  %middle
  \fmffixedy{0}{i2,v3}
  \fmffixedy{0}{v3,v4}
  \fmffixedy{0}{v4,o2}
  \fmf{phantom}{i2,v3}
  \fmf{plain,tension=0.5}{v3,v4}
  \fmf{fermion,label=$d$,l.s=left,l.d=4}{v4,o2}
  \fmf{fermion,tension=0,label=$n$,l.s=left,l.d=3}{v6,v3}
  %bottom
  \fmffixedy{0}{i3,v5}
  \fmffixedy{0}{v5,v6}
  \fmffixedy{0}{v6,o3}
  \fmf{fermion,label=$b$,l.s=left,l.d=4}{i3,v5}
  \fmf{plain,tension=0.5}{v5,v6}
  \fmf{phantom}{v6,o3}
  %vertical
  \fmf{dashes}{v1,v3}
  \fmf{dashes}{v2,v6}
  %diagram label
  \fmfforce{(0.5w,-25)}{c}
  \fmfv{label=$5$,l.d=0}{c}
\end{fmfgraph*}
}
%MBPT 2.6
\fmfframe(10,20)(10,30){
\begin{fmfgraph*}(80,75)
  \fmfstraight
  \fmfpen{thin}
  \fmfset{arrow_len}{3mm}
  %external vertices
  \fmfleft{i4,i3,i2,i1}
  \fmfright{o4,o3,o2,o1}
  %top
  \fmffixedy{0}{i1,v1}
  \fmffixedy{0}{v1,v2}
  \fmffixedy{0}{v2,o1}
  \fmf{fermion,label=$a$,l.s=left,l.d=4}{i1,v1}
  \fmf{plain,tension=0.5}{v1,v2}
  \fmf{phantom}{v2,o1}
  \fmf{fermion,tension=0,label=$m$,l.s=left,l.d=3}{v2,v3}
  %middle
  \fmffixedy{0}{i2,v3}
  \fmffixedy{0}{v3,v4}
  \fmffixedy{0}{v4,o2}
  \fmf{phantom}{i2,v3}
  \fmf{plain,tension=0.5}{v3,v4}
  \fmf{fermion,label=$c$,l.s=left,l.d=4}{v4,o2}
  %bottom
  \fmffixedy{0}{i3,v5}
  \fmffixedy{0}{v5,v6}
  \fmffixedy{0}{v6,o3}
  \fmf{fermion,label=$b$,l.s=left,l.d=4}{i3,v5}
  \fmf{plain,tension=0.5}{v5,v6}
  \fmf{phantom}{v6,o3}
  \fmf{fermion,tension=0,label=$n$,l.s=left,l.d=3}{v6,v7}
  %bottommost
  \fmffixedy{0}{i4,v7}
  \fmffixedy{0}{v7,v8}
  \fmffixedy{0}{v8,o4}
  \fmf{phantom}{i4,v7}
  \fmf{plain,tension=0.5}{v7,v8}
  \fmf{fermion,label=$d$,l.s=left,l.d=4}{v8,o4}
  %vertical
  \fmf{dashes}{v3,v7}
  \fmf{dashes}{v2,v6}
  %diagram label
  \fmfforce{(0.5w,-25)}{c}
  \fmfv{label=$6$,l.d=0}{c}
\end{fmfgraph*}
}
\end{fmffile}
\end{figure}
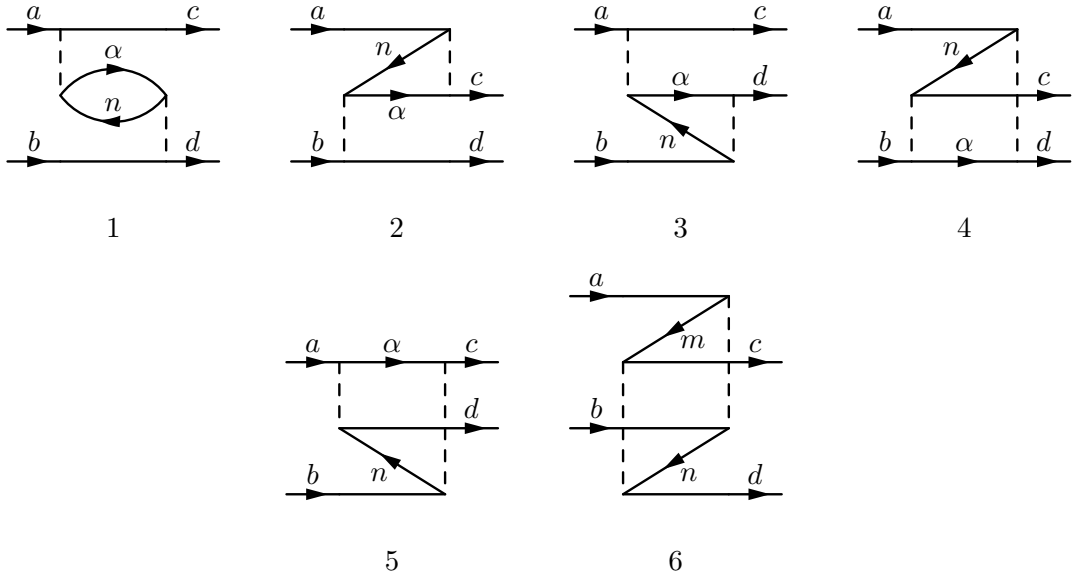

%--------------------------------------------------------------------------------------------

It is worth noting also that the two-body diagrams with the largest contribution to the energy (the direct diagram, \Fig{feyn:sigma2}.1, and its mirror) make no (linear) contribution to the SMS because they cancel. The exchange diagrams do contribute to the SMS; nevertheless, it is because of this cancellation that the contribution of $\Sigma^{(2)}$ to $k_{\rm SMS}$ is generally smaller than that of $\Sigma^{(1)}$.

\begin{figure}[htb]
\caption{Effective three-valence-electron interaction of $\Sigma_2$.}
\begin{fmffile}{sigma3}
\label{feyn:sigma3}
%MBPT 3
\fmfframe(10,10)(10,10){
\begin{fmfgraph}(80,100)
  \fmfstraight
  \fmfpen{thin}
  \fmfset{arrow_len}{3mm}
  %external vertices
  \fmfleft{i4,i3,i2,i1,i0}
  \fmfright{o4,o3,o2,o1,o0}
  %topmost
  \fmffixedy{0}{i0,v0}
  \fmffixedy{0}{v0,v9}
  \fmffixedy{0}{v9,o0}
  \fmf{fermion}{i0,v0}
  \fmf{plain,tension=0.5}{v0,v9}
  \fmf{fermion}{v9,o0}
  %top
  \fmffixedy{0}{i1,v1}
  \fmffixedy{0}{v1,v2}
  \fmffixedy{0}{v2,o1}
  \fmf{fermion}{i1,v1}
  \fmf{plain,tension=0.5}{v1,v2}
  \fmf{phantom}{v2,o1}
  \fmf{fermion,tension=0}{v2,v3}
  %middle
  \fmffixedy{0}{i2,v3}
  \fmffixedy{0}{v3,v4}
  \fmffixedy{0}{v4,o2}
  \fmf{phantom}{i2,v3}
  \fmf{plain,tension=0.5}{v3,v4}
  \fmf{fermion}{v4,o2}
  %bottom
  \fmffixedy{0}{i3,v5}
  \fmffixedy{0}{v5,v6}
  \fmffixedy{0}{v5,o3}
  \fmf{fermion}{i3,v5}
  \fmf{plain,tension=0.5}{v5,v6}
  \fmf{fermion}{v6,o3}
  %vertical
  \fmf{dashes}{v9,v2}
  \fmf{dashes}{v3,v5}
\end{fmfgraph}
}
\end{fmffile}
\end{figure}
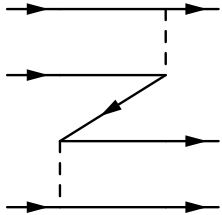

Figure~\ref{feyn:sigma3} shows a diagram with three external lines, $\Sigma^{(3)}$, where three valence electrons interact via the core. The diagrams of this type are easy to calculate (having only one internal summation and no summations over virtual states), but the number of corresponding effective radial integrals is huge. To include this diagram everywhere in the CI calculation would involve taking into account not only corrections to $\mathcal{H}^{(1)}$ and $\mathcal{H}^{(2)}$, but it would also introduce an effective $\mathcal{H}^{(3)}$. Fortunately, as explained in \Cite{dzuba96pra}, it is possible to omit these diagrams entirely provided that one makes an appropriate choice of the atomic core. We have not included $\Sigma^{(3)}$ in this paper.

%--------------------------------------------------------------------------------------------
%--------------------------------------------------------------------------------------------

Analytical expressions for the diagrams of \hbox{Figs. \ref{feyn:sigma1} -- \ref{feyn:sigma2sub}} are given in Appendix~\ref{app:sigma}.

A final point worth mentioning is the definition of the energy denominators in the expressions corresponding to the diagrams. Our formalism corresponds to Brillouin-Wigner perturbation theory, and so there is an explicit dependence on the energy of the whole atom: $E$ in Eqs.~\eref{eq:sigma_def} and \eref{eq:CI_MBPT_matrix}.

From \Eref{eq:PHP}, $E = E_{\rm core} + E_{\rm val}$, and so the core part cancels in the energy denominator $E - \mathcal{H}_0$ (see \Eref{eq:sigma_expansion}). The energy $E_{\rm val}$ corresponds to the energy of all the valence electrons. To calculate $\mathcal{H}_0$, however, one should take into account the state of all valence electrons for each disconnected diagram. This is computationally impractical, but again one can simplify. In this paper all connected diagrams are evaluated at the energies which correspond to the main configuration. For example, when calculating \Fig{feyn:sigma1}.1 the usual Rayleigh-Shr\"odinger perturbation theory would give the denominator ($\epsilon_a + \epsilon_n - \epsilon_\alpha - \epsilon_\beta$); we instead use ($\epsilon_{2s} + \epsilon_n - \epsilon_\alpha - \epsilon_\beta$) if $a$ and $b$ are $s$-wave, or ($\epsilon_{2p_{1/2}} + \epsilon_n - \epsilon_\alpha - \epsilon_\beta$) if $a$ is $p$-wave, and so on.

The Brillouin-Wigner formalism used in our method has two major advantages over the Rayleigh-Shr\"odinger perturbation theory. Firstly, we wish to preserve the symmetry of the CI matrix when $\Sigma$ is added, so we cannot have an energy denominator that depends on which state is initial and which is final. Secondly, Rayleigh-Shr\"odinger theory does not allow a large $P$ space as it will produce small denominators for  excited configurations. That is known to lead to ``intruder states'' -- unphysical states that can lie even below the ground state. By contrast, Brillouin-Wigner theory is known to have the wrong limit for an infinite number of valence electrons. It is possible to formulate the theory without these drawbacks by modifying $E$ (\Cite{kozlov99os}). However, the energy dependence of the effective Hamiltonian is a higher order effect; compared to the prescription above, it leads to a relatively small energy correction that we neglect.

\section{\label{sec:calc} Calculations and results}

We present results for each species of carbon separately, in order of increasing number of valence electrons. For all calculations we use a relativistic B-spline single-electron basis set similar to that developed by Johnson \etal  \cite{johnson86prl,johnson87pra,johnson88pra}. In all cases the MBPT calculation considers $1s^2$ as the core, and all other states as valence. We have used two different B-spline basis sets, each with a different number of shells included in the self-consistent Dirac-Fock procedure. The first set, $\mathcal{B}_1$, is formed in the potential of the closed shell $1s^2$ core; this corresponds to $N_{\rm DF} = N_{\rm core}$. For this set the subtraction diagrams of Figs.~\ref{feyn:sigma1sub} and \ref{feyn:sigma2sub} are zero. The second basis set, $\mathcal{B}_2$, is formed in the potential of the $1s^2 2s^2$ core; thus $N_{\rm DF} > N_{\rm core}$ and the subtraction diagrams must be included.

The basis sets are described by giving the largest principal quantum number for each wave, e.g. $8spd6f$ includes the orbitals $1 - 8s_{1/2}$, $2 - 8p_j$, $3 - 8d_j$, and $4 - 6f_j$. For the MBPT we have used the basis $40spdfg$ for all ions. This basis is fully saturated in the sense that the energies and isotope shifts do not change with the addition of more basis functions.

\subsection{\label{sec:CIV} C\scaps{IV}}

C\scaps{IV} has one electron above a closed $1s^2$ core. It can therefore be treated as a single valence electron atom using MBPT, or as a three electron atom using CI. We have used both methods; the results are presented for comparison in \Tref{tab:CIV}. In each case, the calculations were done using the $\mathcal{B}_1$ basis set. The isotope shift results have also been compared with previous theoretical approaches. Multiconfiguration Hartree-Fock -- CI calculations were presented in \Cite{carlsson95jpb}, which also combined the Hylleraas and full-core plus correlation (FCPC) calculations of Refs.~\cite{king89pra} and \cite{wang93pscr} respectively.

\begin{table}[htbp]
\caption{\label{tab:CIV} Comparison of energies and $k_{\rm SMS}$ transitions from the ground state in C\scaps{IV} using various methods. The transition energies presented do not include the addition of mass shift effects. Note that all results presented by other groups are non-relativistic and hence do not distinguish fine-structure.}
\begin{tabular}{lcccc}
\hline \hline
Level & Energy  & DF + $\Sigma^{(1)}$ & Full CI & Other works \\
      & (expt.) & & & \\
\hline
   & & \multicolumn{3}{c}{Energy (\cm)} \\
$2p\ ^2\!P^o_{1/2}$ & 64484 & 64551 & 64594 & 64564\footnotemark[1] \\
$2p\ ^2\!P^o_{3/2}$ & 64592 & 64681 & 64725 & 64399\footnotemark[2] \\
                    &       &       &       & 64449\footnotemark[3] \\
\\ & & \multicolumn{3}{c}{$k_{\rm SMS}$ (GHz$\cdot$amu)} \\
$2p\ ^2\!P^o_{1/2}$ &  & -4511 & -4521 & -4526\footnotemark[1] \\
$2p\ ^2\!P^o_{3/2}$ &  & -4504 & -4514 & -4527\footnotemark[2] \\
                    &  &       &       & -4528\footnotemark[3] \\
\hline \hline
\end{tabular}
\footnotetext[1]{MCHF--CI: Carlsson \etal, 1995 \cite{carlsson95jpb}}
\footnotetext[2]{MCHF: Godefroid \etal, 2001 \cite{godefroid01jpb}}
\footnotetext[3]{Hylleraas + FCPC: Results of King, 1989 \cite{king89pra}, and Wang \etal, 1993 \cite{wang93pscr}, combined and presented in \Cite{carlsson95jpb}}
\end{table}

As noted previously, the MBPT basis was completely saturated, however we have only included second order diagrams in our calculation. By contrast, the CI calculations are complete (although they do not include the Breit interaction and QED effects), but the basis is not completely saturated. We have used an effective $22spd8f$ basis for the CI calculation, including only single and double excitations from the leading configurations (SD-CI). We included triple excitations for a smaller basis, $14spd8f$: this made a difference of less than 2~\cm~in the transition energy, and less than 1~GHz$\cdot$amu for $k_{\rm SMS}$. Other $f$-wave and higher partial waves were also found to be unimportant.

All methods give the same value for the transition energies and SMS constants to within 0.5\%.

\subsection{\label{sec:CIII} C\scaps{III}}

The ground state for C\scaps{III} is $1s^2 2s^2\ ^1\!S_0$. We have done our calculations both as a four-electron CI problem (full CI) and by combining two-valence-electron CI with MBPT, considering $1s^2$ as the frozen core (CI + $\Sigma^{(1,2)}$). All results are presented in \Tref{tab:CIII}. Also included are CI results (the pure two-electron CI method) and CI + $\Sigma^{(1)}$ results (that do not include $\Sigma^{(2)}$). This allows us to examine the roles of the different parts of the core-correlation. The CI and CI + $\Sigma^{(1)}$ results are calculated with the $\mathcal{B}_2$ basis set; the complete CI + $\Sigma^{(1,2)}$ results have been calculated using both the $\mathcal{B}_1$ and $\mathcal{B}_2$ sets. Additionally we have presented the MCHF results of \Cite{jonsson99jpb}.

\begin{table*}[htbp]
\caption{\label{tab:CIII} Comparison of energies, $k_{\rm SMS}$, and $q$-values for transitions from the ground state in C\scaps{III} using various methods. Note that the MCHF results are non-relativistic and hence do not distinguish fine-structure.}
\begin{tabular}{lccccccc}
\hline \hline
Level & Energy & CI & CI + $\Sigma^{(1)}$ & \multicolumn{2}{c}{CI + $\Sigma^{(1,2)}$}
    & Full CI & MCHF\footnotemark[1]\\
      &(expt.) &    &                     & $\mathcal{B}_2$ & $\mathcal{B}_1$ & & \\
\hline
   & \multicolumn{7}{c}{Energy (\cm)} \\
$2s2p\ ^3\!P^o_0$ & 52367 & 52750 & 52322 & 52349 & 52383 & 52506 & 52280 \\
$2s2p\ ^3\!P^o_1$ & 52391 & 52784 & 52357 & 52383 & 52418 & 52534 & \\
$2s2p\ ^3\!P^o_2$ & 52447 & 52852 & 52427 & 52453 & 52488 & 52592 & \\
$2s2p\ ^1\!P^o_1$ &102352 &103719 &103365 &102725 &102775 &103109 & 102598 \\
\\ & \multicolumn{7}{c}{$k_{\rm SMS}$ (GHz$\cdot$amu)} \\
$2s2p\ ^3\!P^o_0$ &  & -3439 & -3478 & -3473 & -3470 & -3483 & -3475\\
$2s2p\ ^3\!P^o_1$ &  & -3438 & -3476 & -3472 & -3468 & -3480 & \\
$2s2p\ ^3\!P^o_2$ &  & -3434 & -3473 & -3468 & -3465 & -3474 & \\
$2s2p\ ^1\!P^o_1$ &  & -2688 & -2759 & -2790 & -2784 & -2882 & -2817 \\
\\ & \multicolumn{7}{c}{$q$ (\cm)} \\
$2s2p\ ^3\!P^o_0$ &  &  75 &  74 &  74 &  &  & \\
$2s2p\ ^3\!P^o_1$ &  & 109 & 108 & 108 &  &  & \\
$2s2p\ ^3\!P^o_2$ &  & 177 & 178 & 178 &  &  & \\
$2s2p\ ^1\!P^o_1$ &  & 165 & 165 & 165 &  &  & \\
\hline \hline
\end{tabular}
\footnotetext[1]{J\"onsson \etal, 1999 \cite{jonsson99jpb}}
\end{table*}

For the full four-electron CI method we used a very large basis $16spdf$, in the SD-CI approximation. This was not enough to saturate the basis entirely, and we could go no further because the Hamiltonian matrix became too large. The error in $k_{\rm SMS}$ from the full CI calculation could be as large as 100~GHz$\cdot$amu. Nonetheless, they are in agreement with the calculations of \Cite{jonsson99jpb}, as well as the results of our own CI + MBPT.

The CI + MBPT method is particularly accurate for C\scaps{III} for two reasons. Firstly, because there are only two valence electrons there are no triple excitations, which keeps the Hamiltonian small without the need for approximations (for ions with more valence electrons we have used the SD-CI approximation). Because the Hamiltonian is relatively small, we can completely saturate the basis at $20spdf$. Also, there is no $\Sigma^{(3)}$ (corresponding to \Fig{feyn:sigma3}); as stated before, in this paper we have not included it anyway. These points hold true for all two-valence-electron atoms; in particular we have previously shown that excellent results are attainable for Mg\scaps{I} (\Cite{berengut05arXiv}).

In \Cite{jonsson99jpb} the MCHF results were given an error of 1\%; our CI + MBPT results are within this range, and so we believe that we have obtained a similar accuracy. It is also worth noting again that we have excluded the extra box diagrams with ``wrong'' parity from the results presented. The inclusion of these diagrams in $\Sigma^{(2)}$ makes a difference of around 0.1\% to the $k_{\rm SMS}$ constants.

\subsection{\label{sec:CII} C\scaps{II}}

We have treated C\scaps{II} as a three-valence-electron ion; its ground state is $2s^22p\ ^2\!P^o_{1/2}$. We have used the $\mathcal{B}_2$ basis $20spdf$, which corresponds to the $V^{N-1}$ potential, and we have restricted ourselves in the CI problem to single and double excitations from the leading configurations $2s^22p$ and $2s2p^2$.

In \Tref{tab:CII} we present all results for C\scaps{II}. Again we have presented the breakdown of the various parts of the CI + MBPT method. We have also performed our calculations using the $\mathcal{B}_1$ basis: this changed the results by less than 1\% for all results except for the $^2\!S_{1/2}$ transition, in which the difference was around 3\%. For this transition, neither basis set was enough to completely saturate $k_{\rm SMS}$. Furthermore, the difference between the results of CI + MBPT and MCHF--CI is fairly large for this transition (around 7\%). Adding the next most important configuration, $2s^23s$, to the leading set changes the energy of the $^2\!S_{1/2}$ transition by 30~\cm~(0.03\%) and $k_{\rm SMS}$ by 14~GHz$\cdot$amu (around 1\%). The effect on all other transitions was much smaller.

\begin{table*}[htbp]
\caption{\label{tab:CII} Comparison of energies, $k_{\rm SMS}$, and $q$-values for transitions from the ground state in C\scaps{II}. Note that the MCHF--CI results are non-relativistic and hence do not distinguish fine-structure.}
\begin{tabular}{lcccccc}
\hline \hline
Level & Energy  & \multicolumn{3}{c}{This work} & MCHF--CI\footnotemark[1] & CI\footnotemark[2] \\
      & (expt.) & CI & CI + $\Sigma^{(1)}$ & CI + $\Sigma^{(1,2)}$ & & \\
\hline
   & & \multicolumn{5}{c}{Energy (\cm)} \\
$2s2p^2\ ^4\!P_{1/2}$ & 43003  & 43118  &  42767 &  42782 && \\
$2s2p^2\ ^4\!P_{3/2}$ & 43025  & 43144  &  42794 &  42808 && \\
$2s2p^2\ ^4\!P_{5/2}$ & 43054  & 43186  &  42838 &  42852 && \\
$2s2p^2\ ^2\!D_{5/2}$ & 74930  & 75587  &  75350 &  75227 & 74842 & \\
$2s2p^2\ ^2\!D_{3/2}$ & 74933  & 75585  &  75347 &  75225 && \\
$2s2p^2\ ^2\!S_{1/2}$ & 96494  & 97095  &  96965 &  96960 & 96478 & \\
$2s2p^2\ ^2\!P_{1/2}$ & 110624 & 112135 & 111913 & 111205 & 110569 & \\
$2s2p^2\ ^2\!P_{3/2}$ & 110666 & 112187 & 111967 & 111259 && \\
\\ & & \multicolumn{5}{c}{$k_{\rm SMS}$ (GHz$\cdot$amu)} \\
$2s2p^2\ ^4\!P_{1/2}$ &   & -2913 & -2956 & -2960 && \\
$2s2p^2\ ^4\!P_{3/2}$ &   & -2912 & -2954 & -2958 && \\
$2s2p^2\ ^4\!P_{5/2}$ &   & -2910 & -2952 & -2956 && \\
$2s2p^2\ ^2\!D_{5/2}$ &   & -2604 & -2666 & -2672 & -2672 & \\
$2s2p^2\ ^2\!D_{3/2}$ &   & -2604 & -2666 & -2671 && \\
$2s2p^2\ ^2\!S_{1/2}$ &   & -1204 & -1301 & -1321 & -1411 & \\
$2s2p^2\ ^2\!P_{1/2}$ &   & -1323 & -1410 & -1471 & -1531 & \\
$2s2p^2\ ^2\!P_{3/2}$ &   & -1320 & -1407 & -1468 && \\
\\ & & \multicolumn{5}{c}{$q$ (\cm)} \\
$2s2p^2\ ^4\!P_{1/2}$ &   & 132 & 132 & 132 && \\
$2s2p^2\ ^4\!P_{3/2}$ &   & 157 & 158 & 158 && \\
$2s2p^2\ ^4\!P_{5/2}$ &   & 200 & 202 & 202 && \\
$2s2p^2\ ^2\!D_{5/2}$ &   & 179 & 181 & 181 && 179 (20)\\
$2s2p^2\ ^2\!D_{3/2}$ &   & 176 & 178 & 178 && 176 (20)\\
$2s2p^2\ ^2\!S_{1/2}$ &   & 165 & 167 & 168 && 161 (20)\\
$2s2p^2\ ^2\!P_{1/2}$ &   & 162 & 163 & 163 && \\
$2s2p^2\ ^2\!P_{3/2}$ &   & 215 & 217 & 217 && \\
\hline \hline
\end{tabular}
\footnotetext[1]{J\"onsson \etal, 1996 \cite{jonsson96jpb}}
\footnotetext[2]{Berengut \etal, 2004 \cite{berengut04praB}}
\end{table*}

\subsection{\label{sec:CI} C\scaps{I}}

In \Tref{tab:CI_energies} we present energies for transitions in neutral carbon. The ground state of C\scaps{I} is $2s^22p^2\ ^3P_0$. We used the $\mathcal{B}_2$ basis of size $16spdf$, and took all single and double excitations from several leading configurations. The energies obtained for this atom are not as good as those of the other ions. It is testimony to the power of the B-spline basis, however, that the levels appear in the correct order (with the exception of some fine structure), which is remarkable considering that the spectrum is very dense.

\begin{table*}[htbp]
\caption{\label{tab:CI_energies} Comparison of energies and $q$-values for transitions from the ground state in C\scaps{I} using various methods. Note that the MCHF--CI results are non-relativistic and hence do not distinguish fine-structure.}
\begin{tabular}{lccccccc}
\hline \hline
Level & \multicolumn{5}{c}{Energy (\cm)} & \multicolumn{2}{c}{$q$ (\cm)} \\
      & (expt.) & CI & CI + $\Sigma^{(1)}$ & CI + $\Sigma^{(1,2)}$ & MCHF--CI 
      & CI + $\Sigma^{(1,2)}$ & CI\footnotemark[1] \\
\hline
$2s^22p^2\ ^1\!S_0$   & 21648 & 22140 & 22213 & 22335 & 21753\footnotemark[2] &  38 & \\
$2s2p^3\ ^5\!S^o_2$   & 33735 & 33529 & 33234 & 33240 & 33498\footnotemark[2] & 146 & \\
$2s^22p3s\ ^3\!P^o_0$ & 60333 & 59806 & 60182 & 60144 & & -47 & \\
$2s^22p3s\ ^3\!P^o_1$ & 60353 & 59826 & 60202 & 60164 & & -24 & \\
$2s^22p3s\ ^3\!P^o_2$ & 60393 & 59866 & 60243 & 60206 & &  26 & \\
$2s^22p3s\ ^1\!P^o_1$ & 61982 & 61587 & 61975 & 61911 & 62002\footnotemark[2] &   1 &\\
$2s2p^3\ ^3\!D^o_3$   & 64087 & 64773 & 64628 & 64562 & & 144 & 151(60) \\
$2s2p^3\ ^3\!D^o_1$   & 64090 & 64762 & 64617 & 64551 & 64036\footnotemark[3] & 137 & 141(60) \\
$2s2p^3\ ^3\!D^o_2$   & 64091 & 64766 & 64622 & 64555 & & 140 & 145(60) \\
$2s2p^3\ ^3\!P^o_1$   & 75254 & 76209 & 76196 & 76153 & & 117 & 111(60) \\
$2s2p^3\ ^3\!P^o_2$   & 75255 & 76214 & 76202 & 76158 & & 121 & \\
$2s2p^3\ ^3\!P^o_0$   & 75256 & 76207 & 76194 & 76151 & & 115 & \\
$2s^22p3d\ ^1\!D^o_2$ & 77680 & 79297 & 79724 & 79643 & &   7 & \\
$2s^22p4s\ ^3\!P^o_0$ & 78105 & 79737 & 80152 & 80076 & & -33 & \\
$2s^22p4s\ ^3\!P^o_1$ & 78117 & 79750 & 80166 & 80090 & & -21 & \\
$2s^22p4s\ ^3\!P^o_2$ & 78148 & 79787 & 80205 & 80130 & &  24 & \\
$2s^22p3d\ ^3\!F^o_2$ & 78199 & 79845 & 80271 & 80193 & & -31 & \\
$2s^22p3d\ ^3\!F^o_3$ & 78216 & 79862 & 80289 & 80211 & & -18 & \\
%$2s^22p3d\ ^3\!F^o_4$ & 78250
$2s^22p3d\ ^3\!D^o_1$ & 78293 & 79937 & 80354 & 80275 & & -13 & \\
$2s^22p3d\ ^3\!D^o_2$ & 78308 & 79954 & 80371 & 80293 & &  13 & \\
$2s^22p3d\ ^3\!D^o_3$ & 78318 & 79966 & 80385 & 80306 & &  29 & \\
$2s^22p4s\ ^1\!P^o_1$ & 78340 & 79983 & 80403 & 80327 & &  17 & \\
$2s^22p3d\ ^1\!F^o_3$ & 78530 & 80802 & 80626 & 80549 & &  15 & \\
$2s^22p3d\ ^1\!P^o_1$ & 78731 & 80402 & 80822 & 80746 & &  12 & \\
$2s^22p3d\ ^3\!P^o_2$ & 79311 & 80957 & 81307 & 81225 & &  18 & \\
$2s^22p3d\ ^3\!P^o_1$ & 79319 & 80967 & 81318 & 81237 & &  30 & \\
$2s^22p3d\ ^3\!P^o_0$ & 79323 & 80972 & 81323 & 81242 & &  35 & \\
$2s^22p4d\ ^1\!D^o_2$ & 83498 & 86521 & 86942 & 86858 & &   8 & \\
\hline \hline
\end{tabular}
\footnotetext[1]{Berengut \etal, 2004 \cite{berengut04praB}}
\footnotetext[2]{Carlsson \etal, 1995 \cite{carlsson95jpb}}
\footnotetext[3]{J\"onsson \etal, 1996 \cite{jonsson96jpb}}
\end{table*}

We have generated $q$-values for C\scaps{I} because the previous calculations (\Cite{berengut04praB}) had large uncertainties. We believe the new values, presented in \Tref{tab:CI_energies}, have errors of around 10~\cm.

In \Tref{tab:CI_SMS} we present the SMS constants for C\scaps{I}. We are within 1\% of the values obtained using the MCHF--CI method (Refs.~\cite{carlsson95jpb} and \cite{jonsson96jpb}). For most transitions the effect of core-correlations on $k_{\rm SMS}$ is around 1 or 2\%, however in some cases they are larger (for example, in $2s2p^3\ ^3\!P^o$ the core correlations are 8\% of the total).

\begin{table*}[htbp]
\caption{\label{tab:CI_SMS} Comparison of $k_{\rm SMS}$ for transitions from the ground state in C\scaps{I} using various methods. Note that the MCHF--CI results are non-relativistic and hence do not distinguish fine-structure.}
\begin{tabular}{lccccc}
\hline \hline
Level & Energy  & \multicolumn{4}{c}{$k_{\rm SMS}$ (GHz$\cdot$amu)} \\
      & (expt.) & CI & CI + $\Sigma^{(1)}$ & CI + $\Sigma^{(1,2)}$ & MCHF--CI \\
\hline
$2s^22p^2\ ^1\!S_0$   & 21648 &   186 &   180 &   191 & 152\footnotemark[1] \\
$2s2p^3\ ^5\!S^o_2$   & 33735 & -2540 & -2579 & -2588 & -2583\footnotemark[1] \\
$2s^22p3s\ ^3\!P^o_0$ & 60333 &  1405 &  1405 &  1419 & \\
$2s^22p3s\ ^3\!P^o_1$ & 60353 &  1406 &  1406 &  1420 & \\
$2s^22p3s\ ^3\!P^o_2$ & 60393 &  1408 &  1408 &  1422 & \\
$2s^22p3s\ ^1\!P^o_1$ & 61982 &  1549 &  1551 &  1559 & 1553\footnotemark[1] \\
$2s2p^3\ ^3\!D^o_3$   & 64087 & -2165 & -2224 & -2227 & \\
$2s2p^3\ ^3\!D^o_1$   & 64090 & -2165 & -2224 & -2227 & -2222\footnotemark[2] \\
$2s2p^3\ ^3\!D^o_2$   & 64091 & -2165 & -2224 & -2227 & \\
$2s2p^3\ ^3\!P^o_1$   & 75254 & -1272 & -1390 & -1392 & \\
$2s2p^3\ ^3\!P^o_2$   & 75255 & -1272 & -1389 & -1391 & \\
$2s2p^3\ ^3\!P^o_0$   & 75256 & -1271 & -1390 & -1392 & \\
$2s^22p3d\ ^1\!D^o_2$ & 77680 &  1334 &  1320 &  1331 & \\
$2s^22p4s\ ^3\!P^o_0$ & 78105 &  1398 &  1392 &  1407 & \\
$2s^22p4s\ ^3\!P^o_1$ & 78117 &  1404 &  1397 &  1412 & \\
$2s^22p4s\ ^3\!P^o_2$ & 78148 &  1415 &  1408 &  1422 & \\
$2s^22p3d\ ^3\!F^o_2$ & 78199 &  1378 &  1368 &  1381 & \\
$2s^22p3d\ ^3\!F^o_3$ & 78216 &  1381 &  1372 &  1384 & \\
%$2s^22p3d\ ^3\!F^o_4$ & 78250
$2s^22p3d\ ^3\!D^o_1$ & 78293 &  1430 &  1422 &  1434 & \\
$2s^22p3d\ ^3\!D^o_2$ & 78308 &  1429 &  1421 &  1432 & \\
$2s^22p3d\ ^3\!D^o_3$ & 78318 &  1430 &  1420 &  1432 & \\
$2s^22p4s\ ^1\!P^o_1$ & 78340 &  1443 &  1435 &  1446 & \\
$2s^22p3d\ ^1\!F^o_3$ & 78530 &  1451 &  1440 &  1452 & \\
$2s^22p3d\ ^1\!P^o_1$ & 78731 &  1436 &  1426 &  1438 & \\
$2s^22p3d\ ^3\!P^o_2$ & 79311 &   948 &   998 &  1010 & \\
$2s^22p3d\ ^3\!P^o_1$ & 79319 &   956 &  1006 &  1018 & \\
$2s^22p3d\ ^3\!P^o_0$ & 79323 &   960 &  1009 &  1021 & \\
$2s^22p4d\ ^1\!D^o_2$ & 83498 &  1277 &  1258 &  1268 & \\
\hline \hline
\end{tabular}
\footnotetext[1]{Carlsson \etal, 1995 \cite{carlsson95jpb}}
\footnotetext[2]{J\"onsson \etal, 1995 \cite{jonsson96jpb}}
\end{table*}

\section{Conclusions}

In this paper we have presented, in detail, a method for calculating isotope shifts and relativistic shifts in atomic spectra. The method uses the finite-field method to reduce the problem to that of an energy calculation, which is carried out using CI for the valence electrons combined with MBPT for the core correlations. Having previously tested the method in magnesium, we have now applied it to transitions in neutral carbon, C\scaps{II}, C\scaps{III}, and C\scaps{IV}. In all cases we have agreement with previous MCHF and MCHF--CI calculations to within a few percent. In \Tref{tab:experiment} we compare our calculations with the few experiments that exist for carbon ions; in all cases agreement is within around 0.005~\cm, which corresponds to an error in $k_{\rm SMS}$ of around 20~GHz$\cdot$amu.

\begin{table}[htbp]
\caption{\label{tab:experiment} Comparison of calculated $^{13}$C -- $^{12}$C isotope shifts with experiment.}
\begin{tabular}{llccc}
\hline \hline
\multicolumn{2}{c}{Transition}& $\lambda$ & \multicolumn{2}{c}{$\delta \nu^{13, 12}$ (\cm)} \\
Lower Level & Upper Level     &  (\AA)    &  Expt. & This work \\
\hline
C\scaps{I} \\
$2s^22p^2\ ^3\!P_2$ & $2s 2p^3\ ^5\!S_2^o$  & 2967 &  0.670(5)\footnotemark[1] &  0.674\footnotemark[2] \\
$2s^22p^2\ ^1\!S_0$ & $2s^22p3s\ ^1\!P_1^o$ & 2479 & -0.156(3)\footnotemark[3] & -0.151 \\
                    &                       &      & -0.156(2)\footnotemark[4] & \\
C\scaps{II} \\
$2s2p^2\ ^2\!S_{1/2}$ & $2s^23p\ ^2\!P_{3/2}^o$ & 2837 & -0.612(2)\footnotemark[3] & -0.617 \\
$2s2p^2\ ^2\!S_{1/2}$ & $2s^23p\ ^2\!P_{1/2}^o$ & 2838 & -0.623(3)\footnotemark[3] & -0.617 \\
\hline \hline
\end{tabular}
\footnotetext[1]{Bernheim and Kittrell, 1980 \cite{bernheim80sab}}
\footnotetext[2]{Actually, the $2s^22p^2\ ^3\!P_0$ -- $2s 2p^3\ ^5\!S_2^o$ transition was calculated.}
\footnotetext[3]{Burnett, 1950 \cite{burnett50prev}}
\footnotetext[4]{Holmes, 1951 \cite{holmes51osa}}
\end{table}

In \Tref{tab:results} we present total isotope shifts for some important transitions. These transitions can be observed in quasar absorption spectra, and can therefore be used to probe variation of $\alpha$ and isotope abundance evolution. The results are presented both in MHz and km/sec: the latter is the preferred form for use in astronomy.

\begin{table*}[htbp]
\caption{\label{tab:results} Total $^{13}$C -- $^{12}$C and $^{14}$C -- $^{12}$C isotope shifts for important transitions. We believe the errors are of the order of 0.1 GHz.}
\begin{tabular}{llccccc}
\hline \hline
\multicolumn{2}{c}{Transition} & $\lambda$ 
    & \multicolumn{2}{c}{$\delta \nu^{13, 12}$} 
    & \multicolumn{2}{c}{$\delta \nu^{14, 12}$} \\
Ground State & Upper Level & (\AA)& (GHz) & (km/sec)\footnotemark[1]
                                                  & (GHz) & (km/sec)\footnotemark[1] \\
\hline
C\scaps{I} \\
$2s^2 2p^2\ ^3\!P_0$ & $2s^2 2p 3s\ ^3\!P_1^o$
                           & 1657 & -2.75 &  0.46 & -5.09 &  0.84 \\
 & $2s 2p^3   \ ^3\!D_1^o$ & 1560 & 21.10 & -3.29 & 39.12 & -6.10 \\
 & $2s 2p^3   \ ^3\!P_1^o$ & 1329 & 16.91 & -2.25 & 31.34 & -4.17 \\
 & $2s^2 2p 4s\ ^3\!P_1^o$ & 1280 & -0.82 &  0.10 & -1.51 &  0.19 \\
 & $2s^2 2p 3d\ ^3\!D_1^o$ & 1277 & -0.94 &  0.12 & -1.75 &  0.22 \\
 & $2s^2 2p 4s\ ^1\!P_1^o$ & 1276 & -1.01 &  0.13 & -1.88 &  0.24 \\
 & $2s^2 2p 3d\ ^3\!P_1^o$ & 1261 &  1.84 & -0.23 &  3.42 & -0.43 \\
\\ C\scaps{II} \\
$2s^2 2p\ ^2\!P_{1/2}^o$ & $2s 2p^2\ ^2\!D_{3/2}$
                          & 1336 & 25.10 & -3.35 & 46.53 & -6.21 \\
 & $2s 2p^2\ ^2\!D_{5/2}$ & 1336 & 25.10 & -3.35 & 46.54 & -6.21 \\
 & $2s 2p^2\ ^2\!S_{1/2}$ & 1037 & 18.70 & -1.94 & 34.66 & -3.59 \\
\\ C\scaps{III} \\
$2s^2\ ^1\!S_0$ & $2s2p\ ^1\!P_1^o$ & 977 & 28.76 & -2.81 & 53.33 & -5.21 \\
\\ C\scaps{IV} \\
$2s\ ^2\!S_{1/2}$ & $2p\ ^2\!P_{1/2}^o$
                       & 1551 & 35.89 & -5.57 & 66.54 & -10.32 \\
 & $2p\ ^2\!P_{3/2}^o$ & 1548 & 35.79 & -5.54 & 66.35 & -10.27 \\
\hline \hline
\end{tabular}
\footnotetext[1]{$\delta \nu = \delta \lambda / \lambda \times c$ (km/sec).
A negative velocity means that $^{14}$C (and $^{13}$C) are at lower wavelength than $^{12}$C.}
\end{table*}

\section{Acknowledgements}

This work is supported by the Australian Research Council; Department
of Energy, Office of Nuclear Physics, Contract No. W-31-109-ENG-38;
Gordon Godfrey fund; and Russian Foundation for Basic Research,
grant No.~05-02-16914.
The authors would like to thank V. A. Dzuba for useful discussions.
We are grateful to the APAC National Facility for providing computer time.

\appendix
\section{\label{app:sms_operator} Matrix element of the two-body operator}

The two-body operator used is this work is the sum of the Coulomb interaction
operator and the ``rescaled'' SMS operator (atomic units):
\begin{equation}
	\tilde Q = \frac{1}{|{\mathbf r}_1 - {\mathbf r}_2|} + \lambda
	 {\mathbf p}_1 \cdot {\mathbf p}_2 \equiv \sum_k \tilde Q_k,
\label{a1}
\end{equation}
where $\lambda$ is the scaling factor, ${\mathbf p} = -i\nabla$ is electron 
momentum, and
\begin{equation}
	\tilde Q_k = \frac{4\pi}{2k+1}\frac{r_<^k}{r_>^{k+1}}
	Y_{k}({\mathbf n}_1)Y_{k}({\mathbf n}_2) + \lambda\, 
	{\mathbf p}_1 \cdot {\mathbf p}_2\, \delta_{k1}.
\label{a2}
\end{equation}
We use the following form for the single-electron wave function
\begin{eqnarray}
	\phi({\mathbf r})_{jlm} = \frac{1}{r}
 	\left( \begin{array}{c}
	f(r) \Omega({\mathbf n})_{jlm} \\ i 
	\alpha g(r) \tilde{\Omega}({\mathbf n})_{jlm}
	\end{array}
	\right).
\label{psi}
\end{eqnarray}
Here $\alpha = 1/137.036$ is the fine structure constant, and 
$\tilde{\Omega}({\mathbf n})_{jlm} = -(\vec \sigma \cdot {\mathbf n}) 
{\Omega}({\mathbf n})_{jlm}$.

\begin{widetext}
The matrix element of operator (\ref{a2}) with wave functions (\ref{psi})
has the form
\begin{equation}
  \left< \phi_a({\mathbf r}_1)\phi_b({\mathbf r}_2) \right| \tilde Q_k
  \left| \phi_c({\mathbf r}_1)\phi_d({\mathbf r}_2) \right>
 = C^k_{ab,cd} (R^k_{ab,cd} - \lambda \delta_{k1} p_{ac} p_{bd}),
\label{a3}
\end{equation}
where the angular factor $C_k$ is the same for both operators
\begin{eqnarray}
	& C^k_{ab,cd} = (-1)^{q+m_a+m_b} \left( 
	\begin{array}{ccc} j_a & k & j_c \\ -m_a & q & m_c \end{array}
	\right) \left(
	\begin{array}{ccc} j_b & k & j_d \\ -m_b & -q & m_d \end{array} 
	\right) \nonumber \\ 
\label{a4}
	& \times (-1)^{j_a+j_b+j_c+j_d+1}
	\sqrt{(2j_a+1)(2j_b+1)(2j_c+1)(2j_d+1)} \\*
	& \times \left(
	\begin{array}{ccc} j_a & j_c & k \\ \frac{1}{2} & -\frac{1}{2} & 0 
	\end{array}  \right)  \left(
	\begin{array}{ccc} j_b & j_d & k \\ \frac{1}{2} & -\frac{1}{2} & 0 
	\end{array}  \right) \xi(l_a+l_c+k)\xi(l_b+l_d+k) \nonumber, \\*
	& \xi(x) = \left\{ \begin{array}{ll} 1, & \text{if} ~x~ 
	\text{is even} \\*
	0, & \text{if} ~x~ \text{is odd} \end{array} \right. \nonumber,
\end{eqnarray}
$R^k_{ab,cd}$ is radial Coulomb integral
\begin{equation}
R^k_{ab,cd} = \int_0^{\infty}\frac{r_<^k}{r_>^{k+1}}
	\left(f_a(r_1)f_c(r_1)+\alpha^2 g_a(r_1)g_c(r_1)\right)
	\left(f_b(r_2)f_d(r_2)+\alpha^2 g_b(r_2)g_d(r_2)\right) dr_1dr_2 ,
\label{a5}
\end{equation}
while $p_{12}$ is the radial matrix element of the SMS operator
\begin{eqnarray}
\label{a6}
p_{12} &=& A_{12} \delta_{l_1 l_2+1} + B_{12} \delta_{l_1 l_2-1}, \\
A_{12} &=& \int_0^{\infty}f_1(\frac{d}{dr}-\frac{l_1}{r})f_2 dr,	\nonumber \\
B_{12} &=& \int_0^{\infty}f_1(\frac{d}{dr}+\frac{l_2}{r})f_2 dr.	\nonumber
\end{eqnarray}

\section{\label{app:sigma} Effective integrals for $\Sigma$ diagrams}

Here we give expressions for the diagrams in Figs.~\ref{feyn:sigma1} -- \ref{feyn:sigma2sub}. The following conventions for the indices are used: $a$, $b$, $c$, and $d$ correspond to the external lines (valence electrons); $1 \leq m,n \leq N_{\rm core}$; $N_{\rm core} + 1 \leq \alpha,\beta$. Following \cite{lindgren86book} we use the  shorthand notation $[j,k,\ldots] = (2j + 1)(2k+1)\ldots$. Note also that all Wigner three--j symbols come with a parity selection rule, which is not explicitly given. This is because the three--j terms come from the formula
\[
-[l_1,l_2]^{\half} \left( \begin{array}{ccc}l_1 & k & l_2 \\ 0 & 0 & 0 \end{array} \right)
    \sixj{j_1}{j_2}{k}{l_2}{l_1}{k}
 = \threej{j_1}{j_2}{k} \xi(l_1 + l_2 + k) .
\]
Additionally, the subtraction diagrams (Figs.~\ref{feyn:sigma1sub} and \ref{feyn:sigma2sub}) have terms which are matrix elements of $h^{\rm CI}$ (see \Eref{eq:h_CI}). This operator cannot change the angular momentum of the state; for example $\left< a \right| h^{\rm CI} \left| n \right>$ comes with a factor $\delta_{j_a\, j_n} \delta_{l_a\, l_n}$ which is not presented explicitly.

We start with the contributions to the self-energy matrix elements from the diagrams of Figs.~\ref{feyn:sigma1} and \ref{feyn:sigma1sub}. All of these expressions have the following additional terms that we do not write explicitly: $\delta_{j_a\, j_b}$, $\delta_{m_a\, m_b}$, and $\delta_{l_a\, l_b}$.

\Fig{feyn:sigma1}.1:
\begin{equation}
D_{a,b} = \sum_{n\alpha\beta \atop k} \frac{[j_n,j_\alpha,j_\beta]}{[k]}
          \threej{j_a}{j_\beta}{k}^2 \threej{j_n}{j_\alpha}{k}^2
          \frac{R^k_{an,\beta\alpha} R^k_{\beta\alpha,bn}}
               {\epsilon_a + \epsilon_n - \epsilon_\alpha - \epsilon_\beta}
\end{equation}

\Fig{feyn:sigma1}.2:
\begin{eqnarray}
D_{a,b} &=& \sum_{n\alpha\beta \atop k_1 k_2} (-1)^{k_1 + k_2} [j_n,j_\alpha,j_\beta]
    \threej{j_a}{j_\beta}{k_1} \threej{j_n}{j_\alpha}{k_1} \threej{j_\beta}{j_n}{k_2} \nonumber \\
 && \times \threej{j_\alpha}{j_a}{k_2} \sixj{j_a}{j_\beta}{k_2}{j_n}{j_\alpha}{k_1}
    \frac{R^{k_1}_{an,\beta\alpha} R^{k_2}_{\beta\alpha,nb}}
         {\epsilon_a + \epsilon_n - \epsilon_\alpha - \epsilon_\beta}
\end{eqnarray}

\Fig{feyn:sigma1}.3:
\begin{equation}
D_{a,b} = - \sum_{mn\alpha \atop k} \frac{[j_m,j_n,j_\alpha]}{[k]}
          \threej{j_a}{j_m}{k}^2 \threej{j_n}{j_\alpha}{k}^2
          \frac{R^k_{a\alpha,mn} R^k_{mn,b\alpha}}
               {\epsilon_m + \epsilon_n - \epsilon_\alpha - \epsilon_b}
\end{equation}

\Fig{feyn:sigma1}.4:
\begin{eqnarray}
D_{a,b} &=& \sum_{mn\alpha \atop k_1 k_2} (-1)^{k_1 + k_2 + 1} [j_m,j_n,j_\alpha]
    \threej{j_a}{j_m}{k_1} \threej{j_\alpha}{j_n}{k_1} \threej{j_m}{j_\alpha}{k_2} \nonumber \\
 && \times \threej{j_n}{j_a}{k_2} \sixj{j_a}{j_n}{k_2}{j_\alpha}{j_m}{k_1}
    \frac{R^{k_1}_{a\alpha,mn} R^{k_2}_{mn,\alpha b}}
         {\epsilon_m + \epsilon_n - \epsilon_\alpha - \epsilon_b}
\end{eqnarray}

\Fig{feyn:sigma1sub}.1:
\begin{equation}
D_{a,b} = \sum_{n\alpha} [j_n] \frac{R^0_{an,b\alpha} \left< n \right| h^{\rm CI} \left| \alpha \right>}
                                    {\epsilon_n - \epsilon_\alpha}
\end{equation}

\Fig{feyn:sigma1sub}.2:
\begin{equation}
D_{a,b} = \sum_{n\alpha \atop k} (-1)^{2j_a} [j_n] \threej{j_a}{j_n}{k}^2
          \frac{R^k_{a\alpha,nb} \left< n \right| h^{\rm CI} \left| \alpha \right>}
               {\epsilon_n - \epsilon_\alpha}
\end{equation}

\Fig{feyn:sigma1sub}.3:
\begin{equation}
D_{a,b} = - \sum_n \frac{\left< a \right| h^{\rm CI} \left| n \right> \left< n \right| h^{\rm CI} \left| b \right>}
                 {\epsilon_n - \epsilon_a}
\end{equation}

Next we present contributions to the valence-valence screening diagrams due to the diagrams of Figs.~\ref{feyn:sigma2} and \ref{feyn:sigma2sub}. These diagrams have $k$ as an external parameter (not to be summed over). In the CI procedure they are added to $R^k_{ab,cd}$ of \Eref{a3}.

\Fig{feyn:sigma2}.1:
\begin{equation}
R^k_{ab,cd} = \sum_{n\alpha} \frac{[j_\alpha,j_n]}{[k]}
    \threej{j_\alpha}{j_n}{k}^2
    \frac{R^k_{an,c\alpha} R^k_{\alpha b,nd}}
         {\epsilon_\alpha - \epsilon_n}
\end{equation}

\Fig{feyn:sigma2}.2:
\begin{eqnarray}
R^k_{ab,cd} &=& \sum_{n\alpha \atop k_1} (-1)^{k_1 + k} [j_\alpha,j_n]
    \threej{j_a}{j_n}{k_1} \threej{j_\alpha}{j_c}{k_1} \threej{j_n}{j_\alpha}{k} \nonumber \\
 && \times \threej{j_a}{j_c}{k}^{-1} \sixj{j_a}{j_c}{k}{j_\alpha}{j_n}{k_1}
    \frac{R^{k_1}_{a\alpha,nc} R^k_{nb,\alpha d}}{\epsilon_n - \epsilon_\alpha}
\end{eqnarray}

\Fig{feyn:sigma2}.3:
\begin{eqnarray}
R^k_{ab,cd} &=& \sum_{n\alpha \atop k_1} (-1)^{k_1 + k} [j_\alpha,j_n]
    \threej{j_b}{j_n}{k_1} \threej{j_\alpha}{j_d}{k_1} \threej{j_n}{j_\alpha}{k} \nonumber \\
 && \times \threej{j_b}{j_d}{k}^{-1} \sixj{j_b}{j_d}{k}{j_\alpha}{j_n}{k_1}
    \frac{R^{k_1}_{b\alpha,nd} R^k_{na,\alpha c}}{\epsilon_n - \epsilon_\alpha}
\end{eqnarray}

\Fig{feyn:sigma2}.4:
\begin{eqnarray}
R^k_{ab,cd} &=& \sum_{n\alpha \atop k_1 k_2} (-1)^{j_a + j_b + j_c + j_d + j_\alpha + j_n} [j_\alpha,j_n,k]
    \threej{j_a}{j_n}{k_1} \threej{j_\alpha}{j_d}{k_1}  \nonumber \\
 && \times \threej{j_n}{j_c}{k_2} \threej{j_b}{j_\alpha}{k_2}
    \threej{j_a}{j_c}{k}^{-1} \threej{j_b}{j_d}{k}^{-1} \nonumber \\
 && \times \sixj{j_c}{j_a}{k}{k_1}{k_2}{j_n} \sixj{j_b}{j_d}{k}{k_1}{k_2}{j_\alpha}
    \frac{R^{k_1}_{a\alpha,nd} R^{k_2}_{nb,c\alpha}}{\epsilon_n - \epsilon_\alpha}
\end{eqnarray}

\Fig{feyn:sigma2}.5:
\begin{eqnarray}
R^k_{ab,cd} &=& \sum_{n\alpha \atop k_1 k_2} (-1)^{j_a + j_b + j_c + j_d + j_\alpha + j_n} [j_\alpha,j_n,k]
    \threej{j_a}{j_\alpha}{k_1} \threej{j_n}{j_d}{k_1}  \nonumber \\
 && \times \threej{j_\alpha}{j_c}{k_2} \threej{j_b}{j_n}{k_2}
    \threej{j_a}{j_c}{k}^{-1} \threej{j_b}{j_d}{k}^{-1} \nonumber \\
 && \times \sixj{j_c}{j_a}{k}{k_1}{k_2}{j_\alpha} \sixj{j_b}{j_d}{k}{k_1}{k_2}{j_n}
    \frac{R^{k_1}_{an,\alpha d} R^{k_2}_{\alpha b,cn}}{\epsilon_n - \epsilon_\alpha}
\end{eqnarray}

\Fig{feyn:sigma2}.6:
\begin{eqnarray}
R^k_{ab,cd} &=& \sum_{mn \atop k_1 k_2} (-1)^{j_a + j_b + j_c + j_d + j_m + j_n + k + k_1 + k_2 + 1} [j_m,j_n,k]
    \threej{j_a}{j_m}{k_1} \threej{j_b}{j_n}{k_1}  \nonumber \\
 && \times \threej{j_m}{j_c}{k_2} \threej{j_n}{j_d}{k_2}
    \threej{j_a}{j_c}{k}^{-1} \threej{j_b}{j_d}{k}^{-1} \nonumber \\
 && \times \sixj{j_c}{j_a}{k}{k_1}{k_2}{j_m} \sixj{j_d}{j_b}{k}{k_1}{k_2}{j_n}
    \frac{R^{k_1}_{ab,mn} R^{k_2}_{mn,cd}}{\epsilon_m + \epsilon_n - \epsilon_c - \epsilon_d}
\end{eqnarray}

\Fig{feyn:sigma2sub}.1:
\begin{equation}
R^k_{ab,cd} = - \sum_n \frac{R^k_{nb,cd} \left< a \right| h^{\rm CI} \left| n \right>}
                            {\epsilon_n - \epsilon_a}
\end{equation}

\Fig{feyn:sigma2sub}.2:
\begin{equation}
R^k_{ab,cd} = - \sum_n \frac{R^k_{an,cd} \left< b \right| h^{\rm CI} \left| n \right>}
                            {\epsilon_n - \epsilon_b}
\end{equation}

\end{widetext}

\bibliography{references}

\begin{thebibliography}{32}
\expandafter\ifx\csname natexlab\endcsname\relax\def\natexlab#1{#1}\fi
\expandafter\ifx\csname bibnamefont\endcsname\relax
  \def\bibnamefont#1{#1}\fi
\expandafter\ifx\csname bibfnamefont\endcsname\relax
  \def\bibfnamefont#1{#1}\fi
\expandafter\ifx\csname citenamefont\endcsname\relax
  \def\citenamefont#1{#1}\fi
\expandafter\ifx\csname url\endcsname\relax
  \def\url#1{\texttt{#1}}\fi
\expandafter\ifx\csname urlprefix\endcsname\relax\def\urlprefix{URL }\fi
\providecommand{\bibinfo}[2]{#2}
\providecommand{\eprint}[2][]{\url{#2}}

\bibitem[{\citenamefont{Webb et~al.}(1999)\citenamefont{Webb, Flambaum,
  Churchill, Drinkwater, and Barrow}}]{webb99prl}
\bibinfo{author}{\bibfnamefont{J.~K.} \bibnamefont{Webb}},
  \bibinfo{author}{\bibfnamefont{V.~V.} \bibnamefont{Flambaum}},
  \bibinfo{author}{\bibfnamefont{C.~W.} \bibnamefont{Churchill}},
  \bibinfo{author}{\bibfnamefont{M.~J.} \bibnamefont{Drinkwater}},
  \bibnamefont{and} \bibinfo{author}{\bibfnamefont{J.~D.}
  \bibnamefont{Barrow}}, \bibinfo{journal}{Phys. Rev. Lett.}
  \textbf{\bibinfo{volume}{82}}, \bibinfo{pages}{884} (\bibinfo{year}{1999}).

\bibitem[{\citenamefont{Webb et~al.}(2001)\citenamefont{Webb, Murphy, Flambaum,
  Dzuba, Barrow, Churchill, Prochaska, and Wolfe}}]{webb01prl}
\bibinfo{author}{\bibfnamefont{J.~K.} \bibnamefont{Webb}},
  \bibinfo{author}{\bibfnamefont{M.~T.} \bibnamefont{Murphy}},
  \bibinfo{author}{\bibfnamefont{V.~V.} \bibnamefont{Flambaum}},
  \bibinfo{author}{\bibfnamefont{V.~A.} \bibnamefont{Dzuba}},
  \bibinfo{author}{\bibfnamefont{J.~D.} \bibnamefont{Barrow}},
  \bibinfo{author}{\bibfnamefont{C.~W.} \bibnamefont{Churchill}},
  \bibinfo{author}{\bibfnamefont{J.~X.} \bibnamefont{Prochaska}},
  \bibnamefont{and} \bibinfo{author}{\bibfnamefont{A.~M.} \bibnamefont{Wolfe}},
  \bibinfo{journal}{Phys. Rev. Lett.} \textbf{\bibinfo{volume}{87}},
  \bibinfo{pages}{091301} (\bibinfo{year}{2001}).

\bibitem[{\citenamefont{Murphy et~al.}(2001)\citenamefont{Murphy, Webb,
  Flambaum, Dzuba, Churchill, Prochaska, Barrow, and Wolfe}}]{murphy01mnrasA}
\bibinfo{author}{\bibfnamefont{M.~T.} \bibnamefont{Murphy}},
  \bibinfo{author}{\bibfnamefont{J.~K.} \bibnamefont{Webb}},
  \bibinfo{author}{\bibfnamefont{V.~V.} \bibnamefont{Flambaum}},
  \bibinfo{author}{\bibfnamefont{V.~A.} \bibnamefont{Dzuba}},
  \bibinfo{author}{\bibfnamefont{C.~W.} \bibnamefont{Churchill}},
  \bibinfo{author}{\bibfnamefont{J.~X.} \bibnamefont{Prochaska}},
  \bibinfo{author}{\bibfnamefont{J.~D.} \bibnamefont{Barrow}},
  \bibnamefont{and} \bibinfo{author}{\bibfnamefont{A.~M.} \bibnamefont{Wolfe}},
  \bibinfo{journal}{Mon. Not. R. Astron. Soc.} \textbf{\bibinfo{volume}{327}},
  \bibinfo{pages}{1208} (\bibinfo{year}{2001}).

\bibitem[{\citenamefont{Webb et~al.}(2003)\citenamefont{Webb, Murphy, Flambaum,
  and Curran}}]{webb03ass}
\bibinfo{author}{\bibfnamefont{J.~K.} \bibnamefont{Webb}},
  \bibinfo{author}{\bibfnamefont{M.~T.} \bibnamefont{Murphy}},
  \bibinfo{author}{\bibfnamefont{V.~V.} \bibnamefont{Flambaum}},
  \bibnamefont{and} \bibinfo{author}{\bibfnamefont{S.~J.}
  \bibnamefont{Curran}}, \bibinfo{journal}{Astrophys. Space Sci.}
  \textbf{\bibinfo{volume}{283}}, \bibinfo{pages}{565} (\bibinfo{year}{2003}).

\bibitem[{\citenamefont{Quast et~al.}(2004)\citenamefont{Quast, Reimers, and
  Levshakov}}]{quast04aap}
\bibinfo{author}{\bibfnamefont{R.}~\bibnamefont{Quast}},
  \bibinfo{author}{\bibfnamefont{D.}~\bibnamefont{Reimers}}, \bibnamefont{and}
  \bibinfo{author}{\bibfnamefont{S.~A.} \bibnamefont{Levshakov}},
  \bibinfo{journal}{Astron. Astrophys.} \textbf{\bibinfo{volume}{414}},
  \bibinfo{pages}{L7} (\bibinfo{year}{2004}).

\bibitem[{\citenamefont{Srianand et~al.}(2004)\citenamefont{Srianand, Chand,
  Petitjean, and Aracil}}]{srianand04prl}
\bibinfo{author}{\bibfnamefont{R.}~\bibnamefont{Srianand}},
  \bibinfo{author}{\bibfnamefont{H.}~\bibnamefont{Chand}},
  \bibinfo{author}{\bibfnamefont{P.}~\bibnamefont{Petitjean}},
  \bibnamefont{and} \bibinfo{author}{\bibfnamefont{B.}~\bibnamefont{Aracil}},
  \bibinfo{journal}{Phys. Rev. Lett.} \textbf{\bibinfo{volume}{92}},
  \bibinfo{pages}{121302} (\bibinfo{year}{2004}).

\bibitem[{\citenamefont{Levshakov et~al.}(2005)\citenamefont{Levshakov,
  Centuri\'on, Molaro, and D'Odorico}}]{levshakov05aap}
\bibinfo{author}{\bibfnamefont{S.~A.} \bibnamefont{Levshakov}},
  \bibinfo{author}{\bibfnamefont{M.}~\bibnamefont{Centuri\'on}},
  \bibinfo{author}{\bibfnamefont{P.}~\bibnamefont{Molaro}}, \bibnamefont{and}
  \bibinfo{author}{\bibfnamefont{S.}~\bibnamefont{D'Odorico}},
  \bibinfo{journal}{Astron. Astrophys.} \textbf{\bibinfo{volume}{434}},
  \bibinfo{pages}{827} (\bibinfo{year}{2005}).

\bibitem[{\citenamefont{Dzuba et~al.}(1999)\citenamefont{Dzuba, Flambaum, and
  Webb}}]{dzuba99prl}
\bibinfo{author}{\bibfnamefont{V.~A.} \bibnamefont{Dzuba}},
  \bibinfo{author}{\bibfnamefont{V.~V.} \bibnamefont{Flambaum}},
  \bibnamefont{and} \bibinfo{author}{\bibfnamefont{J.~K.} \bibnamefont{Webb}},
  \bibinfo{journal}{Phys. Rev. Lett.} \textbf{\bibinfo{volume}{82}},
  \bibinfo{pages}{888} (\bibinfo{year}{1999}).

\bibitem[{\citenamefont{Kozlov et~al.}(2004)\citenamefont{Kozlov, Korol,
  Berengut, Dzuba, and Flambaum}}]{kozlov04pra}
\bibinfo{author}{\bibfnamefont{M.~G.} \bibnamefont{Kozlov}},
  \bibinfo{author}{\bibfnamefont{V.~A.} \bibnamefont{Korol}},
  \bibinfo{author}{\bibfnamefont{J.~C.} \bibnamefont{Berengut}},
  \bibinfo{author}{\bibfnamefont{V.~A.} \bibnamefont{Dzuba}}, \bibnamefont{and}
  \bibinfo{author}{\bibfnamefont{V.~V.} \bibnamefont{Flambaum}},
  \bibinfo{journal}{Phys. Rev. A} \textbf{\bibinfo{volume}{70}},
  \bibinfo{pages}{062108} (\bibinfo{year}{2004}).

\bibitem[{\citenamefont{Ashenfelter
  et~al.}(2004{\natexlab{a}})\citenamefont{Ashenfelter, Mathews, and
  Olive}}]{ashenfelter04prl}
\bibinfo{author}{\bibfnamefont{T.~P.} \bibnamefont{Ashenfelter}},
  \bibinfo{author}{\bibfnamefont{G.~J.} \bibnamefont{Mathews}},
  \bibnamefont{and} \bibinfo{author}{\bibfnamefont{K.~A.} \bibnamefont{Olive}},
  \bibinfo{journal}{Phys. Rev. Lett.} \textbf{\bibinfo{volume}{92}},
  \bibinfo{pages}{041102} (\bibinfo{year}{2004}{\natexlab{a}}).

\bibitem[{\citenamefont{Ashenfelter
  et~al.}(2004{\natexlab{b}})\citenamefont{Ashenfelter, Mathews, and
  Olive}}]{ashenfelter04apj}
\bibinfo{author}{\bibfnamefont{T.~P.} \bibnamefont{Ashenfelter}},
  \bibinfo{author}{\bibfnamefont{G.~J.} \bibnamefont{Mathews}},
  \bibnamefont{and} \bibinfo{author}{\bibfnamefont{K.~A.} \bibnamefont{Olive}},
  \bibinfo{journal}{Astrophys. J.} \textbf{\bibinfo{volume}{615}},
  \bibinfo{pages}{82} (\bibinfo{year}{2004}{\natexlab{b}}).

\bibitem[{\citenamefont{Fenner et~al.}(2005)\citenamefont{Fenner, Murphy, and
  Gibson}}]{fenner05mnras}
\bibinfo{author}{\bibfnamefont{Y.}~\bibnamefont{Fenner}},
  \bibinfo{author}{\bibfnamefont{M.~T.} \bibnamefont{Murphy}},
  \bibnamefont{and} \bibinfo{author}{\bibfnamefont{B.~K.}
  \bibnamefont{Gibson}}, \bibinfo{journal}{Mon. Not. R. Astron. Soc.}
  \textbf{\bibinfo{volume}{358}}, \bibinfo{pages}{468} (\bibinfo{year}{2005}).

\bibitem[{\citenamefont{Carlsson et~al.}(1995)\citenamefont{Carlsson,
  J\"{o}nsson, Godefroid, and \hbox{Froese~Fischer}}}]{carlsson95jpb}
\bibinfo{author}{\bibfnamefont{J.}~\bibnamefont{Carlsson}},
  \bibinfo{author}{\bibfnamefont{P.}~\bibnamefont{J\"{o}nsson}},
  \bibinfo{author}{\bibfnamefont{M.~R.} \bibnamefont{Godefroid}},
  \bibnamefont{and}
  \bibinfo{author}{\bibfnamefont{C.}~\bibnamefont{\hbox{Froese~Fischer}}},
  \bibinfo{journal}{J. Phys. B: At. Mol. Opt. Phys.}
  \textbf{\bibinfo{volume}{28}}, \bibinfo{pages}{3729} (\bibinfo{year}{1995}).

\bibitem[{\citenamefont{Godefroid et~al.}(2001)\citenamefont{Godefroid,
  \hbox{Froese~Fischer}, and J\"{o}nsson}}]{godefroid01jpb}
\bibinfo{author}{\bibfnamefont{M.}~\bibnamefont{Godefroid}},
  \bibinfo{author}{\bibfnamefont{C.}~\bibnamefont{\hbox{Froese~Fischer}}},
  \bibnamefont{and}
  \bibinfo{author}{\bibfnamefont{P.}~\bibnamefont{J\"{o}nsson}},
  \bibinfo{journal}{J. Phys. B: At. Mol. Opt. Phys.}
  \textbf{\bibinfo{volume}{34}}, \bibinfo{pages}{1079} (\bibinfo{year}{2001}).

\bibitem[{\citenamefont{J\"{o}nsson et~al.}(1999)\citenamefont{J\"{o}nsson,
  \hbox{Froese~Fischer}, and Godefroid}}]{jonsson99jpb}
\bibinfo{author}{\bibfnamefont{P.}~\bibnamefont{J\"{o}nsson}},
  \bibinfo{author}{\bibfnamefont{C.}~\bibnamefont{\hbox{Froese~Fischer}}},
  \bibnamefont{and} \bibinfo{author}{\bibfnamefont{M.~R.}
  \bibnamefont{Godefroid}}, \bibinfo{journal}{J. Phys. B: At. Mol. Opt. Phys.}
  \textbf{\bibinfo{volume}{32}}, \bibinfo{pages}{1233} (\bibinfo{year}{1999}).

\bibitem[{\citenamefont{J\"{o}nsson et~al.}(1996)\citenamefont{J\"{o}nsson,
  \hbox{Froese~Fischer}, and Godefroid}}]{jonsson96jpb}
\bibinfo{author}{\bibfnamefont{P.}~\bibnamefont{J\"{o}nsson}},
  \bibinfo{author}{\bibfnamefont{C.}~\bibnamefont{\hbox{Froese~Fischer}}},
  \bibnamefont{and} \bibinfo{author}{\bibfnamefont{M.~R.}
  \bibnamefont{Godefroid}}, \bibinfo{journal}{J. Phys. B: At. Mol. Opt. Phys.}
  \textbf{\bibinfo{volume}{29}}, \bibinfo{pages}{2393} (\bibinfo{year}{1996}).

\bibitem[{\citenamefont{Dzuba et~al.}(1996)\citenamefont{Dzuba, Flambaum, and
  Kozlov}}]{dzuba96pra}
\bibinfo{author}{\bibfnamefont{V.~A.} \bibnamefont{Dzuba}},
  \bibinfo{author}{\bibfnamefont{V.~V.} \bibnamefont{Flambaum}},
  \bibnamefont{and} \bibinfo{author}{\bibfnamefont{M.~G.}
  \bibnamefont{Kozlov}}, \bibinfo{journal}{Phys. Rev. A}
  \textbf{\bibinfo{volume}{54}}, \bibinfo{pages}{3948} (\bibinfo{year}{1996}).

\bibitem[{\citenamefont{Dzuba and Johnson}(1998)}]{dzuba98pra}
\bibinfo{author}{\bibfnamefont{V.~A.} \bibnamefont{Dzuba}} \bibnamefont{and}
  \bibinfo{author}{\bibfnamefont{W.~R.} \bibnamefont{Johnson}},
  \bibinfo{journal}{Phys. Rev. A} \textbf{\bibinfo{volume}{57}},
  \bibinfo{pages}{2459} (\bibinfo{year}{1998}).

\bibitem[{\citenamefont{Berengut et~al.}(2005)\citenamefont{Berengut, Flambaum,
  and Kozlov}}]{berengut05arXiv}
\bibinfo{author}{\bibfnamefont{J.~C.} \bibnamefont{Berengut}},
  \bibinfo{author}{\bibfnamefont{V.~V.} \bibnamefont{Flambaum}},
  \bibnamefont{and} \bibinfo{author}{\bibfnamefont{M.~G.} \bibnamefont{Kozlov}}
  (\bibinfo{year}{2005}), \eprint{arXiv:physics/0507062}.

\bibitem[{\citenamefont{Sobel'man}(1979)}]{sobelman79book}
\bibinfo{author}{\bibfnamefont{I.~I.} \bibnamefont{Sobel'man}},
  \emph{\bibinfo{title}{Atomic Spectra and Radiative Transitions}}
  (\bibinfo{publisher}{Springer-Verlag}, \bibinfo{address}{Berlin},
  \bibinfo{year}{1979}).

\bibitem[{\citenamefont{Berengut et~al.}(2003)\citenamefont{Berengut, Dzuba,
  and Flambaum}}]{berengut03pra}
\bibinfo{author}{\bibfnamefont{J.~C.} \bibnamefont{Berengut}},
  \bibinfo{author}{\bibfnamefont{V.~A.} \bibnamefont{Dzuba}}, \bibnamefont{and}
  \bibinfo{author}{\bibfnamefont{V.~V.} \bibnamefont{Flambaum}},
  \bibinfo{journal}{Phys. Rev. A} \textbf{\bibinfo{volume}{68}},
  \bibinfo{pages}{022502} (\bibinfo{year}{2003}).

\bibitem[{\citenamefont{Lindgren and Morrison}(1986)}]{lindgren86book}
\bibinfo{author}{\bibfnamefont{I.}~\bibnamefont{Lindgren}} \bibnamefont{and}
  \bibinfo{author}{\bibfnamefont{J.}~\bibnamefont{Morrison}},
  \emph{\bibinfo{title}{Atomic Many-Body Theory}}
  (\bibinfo{publisher}{Springer-Verlag}, \bibinfo{address}{New York},
  \bibinfo{year}{1986}).

\bibitem[{\citenamefont{Kozlov and Porsev}(1999)}]{kozlov99os}
\bibinfo{author}{\bibfnamefont{M.~G.} \bibnamefont{Kozlov}} \bibnamefont{and}
  \bibinfo{author}{\bibfnamefont{S.~G.} \bibnamefont{Porsev}},
  \bibinfo{journal}{Opt. Spectrosc.} \textbf{\bibinfo{volume}{87}},
  \bibinfo{pages}{352} (\bibinfo{year}{1999}).

\bibitem[{\citenamefont{Johnson and Sapirstein}(1986)}]{johnson86prl}
\bibinfo{author}{\bibfnamefont{W.~R.} \bibnamefont{Johnson}} \bibnamefont{and}
  \bibinfo{author}{\bibfnamefont{J.}~\bibnamefont{Sapirstein}},
  \bibinfo{journal}{Phys. Rev. Lett.} \textbf{\bibinfo{volume}{57}},
  \bibinfo{pages}{1126} (\bibinfo{year}{1986}).

\bibitem[{\citenamefont{Johnson et~al.}(1987)\citenamefont{Johnson, Idrees, and
  Sapirstein}}]{johnson87pra}
\bibinfo{author}{\bibfnamefont{W.~R.} \bibnamefont{Johnson}},
  \bibinfo{author}{\bibfnamefont{M.}~\bibnamefont{Idrees}}, \bibnamefont{and}
  \bibinfo{author}{\bibfnamefont{J.}~\bibnamefont{Sapirstein}},
  \bibinfo{journal}{Phys. Rev. A} \textbf{\bibinfo{volume}{35}},
  \bibinfo{pages}{3218} (\bibinfo{year}{1987}).

\bibitem[{\citenamefont{Johnson et~al.}(1988)\citenamefont{Johnson, Blundell,
  and Sapirstein}}]{johnson88pra}
\bibinfo{author}{\bibfnamefont{W.~R.} \bibnamefont{Johnson}},
  \bibinfo{author}{\bibfnamefont{S.~A.} \bibnamefont{Blundell}},
  \bibnamefont{and}
  \bibinfo{author}{\bibfnamefont{J.}~\bibnamefont{Sapirstein}},
  \bibinfo{journal}{Phys. Rev. A.} \textbf{\bibinfo{volume}{37}},
  \bibinfo{pages}{307} (\bibinfo{year}{1988}).

\bibitem[{\citenamefont{King}(1989)}]{king89pra}
\bibinfo{author}{\bibfnamefont{F.~W.} \bibnamefont{King}},
  \bibinfo{journal}{Phys. Rev. A} \textbf{\bibinfo{volume}{40}},
  \bibinfo{pages}{1735} (\bibinfo{year}{1989}).

\bibitem[{\citenamefont{Wang et~al.}(1993)\citenamefont{Wang, Zhu, and
  Chung}}]{wang93pscr}
\bibinfo{author}{\bibfnamefont{Z.-W.} \bibnamefont{Wang}},
  \bibinfo{author}{\bibfnamefont{X.-W.} \bibnamefont{Zhu}}, \bibnamefont{and}
  \bibinfo{author}{\bibfnamefont{K.~T.} \bibnamefont{Chung}},
  \bibinfo{journal}{Phys. Scr.} \textbf{\bibinfo{volume}{47}},
  \bibinfo{pages}{65} (\bibinfo{year}{1993}).

\bibitem[{\citenamefont{Berengut et~al.}(2004)\citenamefont{Berengut, Dzuba,
  Flambaum, and Marchenko}}]{berengut04praB}
\bibinfo{author}{\bibfnamefont{J.~C.} \bibnamefont{Berengut}},
  \bibinfo{author}{\bibfnamefont{V.~A.} \bibnamefont{Dzuba}},
  \bibinfo{author}{\bibfnamefont{V.~V.} \bibnamefont{Flambaum}},
  \bibnamefont{and} \bibinfo{author}{\bibfnamefont{M.~V.}
  \bibnamefont{Marchenko}}, \bibinfo{journal}{Phys. Rev. A}
  \textbf{\bibinfo{volume}{70}}, \bibinfo{pages}{064101}
  (\bibinfo{year}{2004}).

\bibitem[{\citenamefont{Bernheim and Kittrell}(1980)}]{bernheim80sab}
\bibinfo{author}{\bibfnamefont{R.~A.} \bibnamefont{Bernheim}} \bibnamefont{and}
  \bibinfo{author}{\bibfnamefont{C.}~\bibnamefont{Kittrell}},
  \bibinfo{journal}{Spectrochim. Acta B} \textbf{\bibinfo{volume}{35}},
  \bibinfo{pages}{51} (\bibinfo{year}{1980}).

\bibitem[{\citenamefont{Burnett}(1950)}]{burnett50prev}
\bibinfo{author}{\bibfnamefont{C.~R.} \bibnamefont{Burnett}},
  \bibinfo{journal}{Phys. Rev.} \textbf{\bibinfo{volume}{80}},
  \bibinfo{pages}{494} (\bibinfo{year}{1950}).

\bibitem[{\citenamefont{Holmes}(1951)}]{holmes51osa}
\bibinfo{author}{\bibfnamefont{J.~R.} \bibnamefont{Holmes}},
  \bibinfo{journal}{J. Opt. Soc. Am.} \textbf{\bibinfo{volume}{41}},
  \bibinfo{pages}{360} (\bibinfo{year}{1951}).

\end{thebibliography}

\end{document}